\documentclass[manuscript]{emulateapj}

 \newcommand{\mum}{${\rm \mu m}$}

 \newcommand{\ks}{ks}

 \newcommand{\nustar}{{\it NuSTAR}}
 \newcommand{\chandra}{{\it Chandra}}
 \newcommand{\xmm}{{\it XMM-Newton}}
 \newcommand{\swift}{{\it Swift}}

\newcommand{\fig}[1]{Figure \ref{#1}}

\newcommand{\Fig}[1]{Figure \ref{#1}}

\def\simgt{\lower.5ex\hbox{\gtsima}}
\def\simlt{\lower.5ex\hbox{\ltsima}}

\shorttitle{NuSTAR ECDFS Survey}
\shortauthors{Mullaney et al.}
\begin{document}

\title{The NuSTAR Extragalactic Surveys: Initial results and catalog from the
  Extended Chandra Deep Field South}

\author{J. R. Mullaney\altaffilmark{1,2},
  A. Del-Moro\altaffilmark{1},
  J. Aird\altaffilmark{1,3}, 
  D. M. Alexander\altaffilmark{1},
  F. M. Civano\altaffilmark{4,5,6},
  R. C. Hickox\altaffilmark{6},
  G. B. Lansbury\altaffilmark{1},
  M. Ajello\altaffilmark{7},
  R. Assef\altaffilmark{8},
  D. R. Ballantyne\altaffilmark{9},
  M. Balokovi\'c\altaffilmark{10},
  F. E. Bauer\altaffilmark{11,12,13},
  W. N. Brandt\altaffilmark{14,15,16},
  S. E. Boggs\altaffilmark{7}, 
  M. Brightman\altaffilmark{10},
  F. E. Christensen\altaffilmark{17},
  A. Comastri\altaffilmark{18}, 
  W. W. Craig\altaffilmark{7},
  M. Elvis\altaffilmark{19},
  K. Forster\altaffilmark{10},
  P. Gandhi\altaffilmark{20,1},
  B. W. Grefenstette\altaffilmark{10},
  C. J. Hailey\altaffilmark{21},
  F. A. Harrison\altaffilmark{10},
  M. Koss\altaffilmark{22},
  S. M. LaMassa\altaffilmark{4},
  B. Luo\altaffilmark{14,15},
  K. K. Madsen\altaffilmark{10},
  S. Puccetti\altaffilmark{23,24},
  C. Saez\altaffilmark{25},
  D. Stern\altaffilmark{26},
  E. Treister\altaffilmark{27},
  C. M. Urry\altaffilmark{4},
  D. R. Wik\altaffilmark{28,29},
  L. Zappacosta\altaffilmark{24}, 
  W. Zhang\altaffilmark{30}
}

\altaffiltext{1}{Centre of Extragalactic Astronomy, Department of Physics, Durham University,
South Road, Durham DH1 3LE, U.K.}
\altaffiltext{2}{The Department of Physics and Astronomy, The University of Sheffield,
Hounsfield Road, Sheffield S3 7RH, U.K.}
\altaffiltext{3}{Institute of Astronomy, University of Cambridge, Madingley Road, Cambridge CB3 0HA}
\altaffiltext{4}{Yale Center for Astronomy and Astrophysics, 260 Whitney ave, New Haven, CT 06520, USA}
\altaffiltext{5}{Smithsonian Astrophysical Observatory, 60 Garden St, Cambridge, MA 02138, USA}
\altaffiltext{6}{Department of Physics and Astronomy, Dartmouth College, 6127 Wilder Laboratory, Hanover, NH 03755, USA}
\altaffiltext{7}{Space Sciences Laboratory, 7 Gauss Way, University of California, Berkeley, CA 94720-7450, USA}
\altaffiltext{8}{N\'ucleo de Astronom\'ia de la Facultad de Ingenier\'ia, Universidad Diego Portales, Av. Ej\'ercito Libertador 441, Santiago, Chile}
\altaffiltext{9}{Center for Relativistic Astrophysics, School of Physics, Georgia Institute of Technology, Atlanta, GA 30332, USA}
\altaffiltext{10}{Cahill Center for Astrophysics, 1216 E. California
  Blvd, California Institute of Technology, Pasadena, CA 91125, USA}
\altaffiltext{11}{Instituto de Astrof\'{\i}sica, Facultad de F\'{i}sica, Pontificia Universidad Cat\'{o}lica de Chile, 306, Santiago 22, Chile} 
\altaffiltext{12}{Millennium Institute of Astrophysics, Santiago, Chile} 
\altaffiltext{13}{Space Science Institute, 4750 Walnut Street, Suite 205, Boulder, Colorado 80301}
\altaffiltext{14}{Department of Astronomy and Astrophysics, The Pennsylvania State University, 525 Davey Lab, University Park, PA 16802, USA}
\altaffiltext{15}{Institute for Gravitation and the Cosmos, The
  Pennsylvania State University, University Park, PA 16802, USA}
\altaffiltext{16}{Department of Physics, The Pennsylvania State University,
  University Park, PA 16802, USA}
\altaffiltext{17}{DTU Space, National Space Institute, Technical University
of Denmark, Elektrovej 327, DK-2800 Lyngby, Denmark}
\altaffiltext{18}{INAF Osservatorio Astronomico di Bologna, via
  Ranzani 1, I-40127, Bologna, Italy}
\altaffiltext{19}{Harvard Smithsonian Center for Astrophysics, 60 Garden St., Cambridge, MA 02138, USA}
\altaffiltext{20}{School of Physics \& Astronomy, University of
  Southampton, Highfield, Southampton SO17 1BJ, UK}
\altaffiltext{21}{Columbia Astrophysics Laboratory, Columbia University,
New York, NY 10027, USA}
\altaffiltext{22}{Institute for Astronomy, Department of Physics, ETH Zurich, Wolfgang-Pauli-Strasse 27, CH-8093 Zurich, Switzerland}
\altaffiltext{23}{ASDC-ASI, Via del Politecnico, I-00133 Roma, Italy}
\altaffiltext{24}{Osservatorio Astronomico di Roma (INAF), via
  Frascati 33, 00040 Monte Porzio Catone (Roma), Italy}
\altaffiltext{25}{Department of Astronomy, University of Maryland, College Park, MD 20742-2421, USA}
\altaffiltext{26}{Jet Propulsion Laboratory, California Institute of Technology, 4800 Oak Grove Drive, Mail Stop 169-221, Pasadena, CA 91109, USA}
\altaffiltext{27}{Universidad de Concepci\'on, Departamento de Astronom\'ia, Casilla 160-C, Concepci\'on, Chile}
\altaffiltext{28}{NASA Goddard Space Flight Center, Code 662,
  Greenbelt, MD 20771, USA}
\altaffiltext{29}{The Johns Hopkins University, Homewood Campus, Baltimore, MD 21218, USA}
\altaffiltext{30}{Physics \& Engineering Department, West Virginia Wesleyan
College, Buckhannon, WV 26201, USA}

\begin{abstract} 
  We present initial results and the source catalog from the \nustar\
  survey of the Extended Chandra Deep Field South (hereafter, ECDFS)
  -- currently the deepest contiguous component of the \nustar\
  extragalactic survey program.  The survey covers the full
  $\approx$30\arcmin$\times$30\arcmin\ area of this field to a maximum
  depth of $\approx$360~ks ($\approx220$~ks when corrected for
  vignetting at 3-24~keV), reaching sensitivity limits of
  $\approx1.3\times10^{-14}~{\rm ergs~s^{-1}~cm^{-2}}$ (3-8~keV),
  $\approx3.4\times10^{-14}~{\rm ergs~s^{-1}~cm^{-2}}$ (8-24~keV) and
  $\approx3.0\times10^{-14}~{\rm ergs~s^{-1}~cm^{-2}}$ (3-24~keV).
  Fifty four (54) sources are detected over the full field, although
  five of these are found to lie below our significance threshold once
  contaminating flux from neighboring (i.e., blended) sources is
  taken into account.  Of the remaining 49 that are significant, 19
  are detected in the 8-24~keV band.  The 8-24~keV to 3-8~keV band
  ratios of the twelve sources that are detected in both bands span
  the range 0.39--1.7, corresponding to a photon index range of
  $\Gamma\approx0.5-2.3$, with a median photon index of
  $\overline{\Gamma}=1.70\pm0.52$.  The redshifts of the 49 sources in
  our main sample span the range $z=0.21-2.7$, and their rest-frame
  10-40~keV luminosities (derived from the observed 8-24~keV fluxes)
  span the range $L_{\rm 10-40keV}\approx(0.7-300)\times10^{43}~{\rm
    ergs~s^{-1}}$, sampling below the ``knee'' of the X-ray luminosity
  function out to $z\sim0.8-1$.  Finally, we identify one \nustar\ source
  that has neither a \chandra\ nor an \xmm\ counterpart, but that
  shows evidence of nuclear activity at infrared wavelengths, and thus
  may represent a genuine, new X-ray source detected by \nustar\ in
  the ECDFS.

\end{abstract}


\section{Introduction}
\label{Introduction}
Extragalactic X-ray surveys have revolutionized our understanding of
the accretion of matter on to supermassive black holes.  Collectively
they have provided a census of active galactic nuclei (AGN) across
broad swathes of parameter space, enabling astronomers to relate black
hole growth to fundamental properties such as large-scale environment
and host galaxy characteristics (e.g., stellar mass, age, morphology;
see the reviews of \citealt{Alexander12} and \citealt{Brandt15}).
Indeed, the deepest \chandra\ and \xmm\ surveys identify significant
populations (i.e., $\sim$tens) of AGNs beyond $z \sim 3$ (e.g.,
\citealt{Alexander03, Luo08, Hasinger08, Elvis09, Civano12, Brusa09,
  Laird09, Xue11, Goulding12, Ranalli13}), probing the history of
black hole growth over $\gtrsim 80\%$ of the age of the Universe.
Such surveys have resolved almost all ($\approx$70--90\%) of the
Cosmic X-ray Background (hereafter, CXB; \citealt{Giacconi62}) at
0.5-8~keV (e.g., \citealt{Worsley05, Hickox06, Lehmer12, Xue12}).

Despite their undeniable success in identifying and characterizing the
AGN population to high redshifts, \chandra\ and \xmm\ are only
sensitive to observed-frame photon energies $\lesssim$10~keV.  This
represents a significant limitation as it is known from earlier,
non-focussing, X-ray missions that the CXB peaks at
$\approx$20--40~keV (e.g., \citealt{Marshall80, Gruber99, Frontera07,
  Churazov07, Ajello08}).  Until recently, the most advanced X-ray
telescopes sensitive to these energies have resolved only 1--2\% of
this peak into individual sources (e.g., \citealt{Krivonos07,
  Ajello08, Bottacini12}).  As such, the sources that make up the peak
of the CXB are almost wholly unconstrained by direct observations,
with the best constraints instead coming from population synthesis
models (i.e., using spectral models to extrapolate the X-ray spectra
of the source populations detected at lower energies; e.g.,
\citealt{Setti89, Madau94, Comastri95, Gilli01, Treister05, Gilli07,
  Treister09, Ballantyne11}).  Many such models require a significant
population of Compton-thick AGNs (i.e., obscured by absorbing column
densities, $N_{\rm H}>1.5\times10^{24}~{\rm cm^{-2}}$, the inverse of
the Thomson cross section) to reproduce the peak of the CXB.  However,
these predictions are heavily influenced by the spectral assumptions
used to extrapolate the X-ray spectra to $>10$~keV, and strong
degeneracies exist between the assumed model parameters (e.g.,
\citealt{Ballantyne06, Gilli07, Treister09, Akylas12}).

Recent progress in characterizing the hard X-ray output from the AGN
population has been made by studies exploiting data collected by the
{\it INTEGRAL} and {\it Swift} telescopes (e.g. \citealt{Krivonos07,
  Tueller08, Ajello08, Burlon11, Ajello12, Vasudevan13}).  These
studies report that $\sim5-25\%$ of local AGNs are confirmed on the
basis of X-ray spectral analyses to be Compton-thick (to $N_{\rm
  H}\sim10^{25}~{\rm cm^{-2}}$).  However, the limited sensitivity of
these telescopes means that they can only probe the most local
($\lesssim100$~Mpc) AGNs to the depth needed to verify Compton-thick
levels of absorption, leaving no constraints on the evolution of the
absorbed fractions of AGNs.  The {\it Nuclear Spectroscopic Telescope
  Array} (hereafter, \nustar; \citealt{Harrison13}) is a factor of
$\sim10^2$ more sensitive than previous high energy X-ray telescopes
and is predicted to determine the make up of 25-35\% of the CXB
(\citealt{Ballantyne11}), allowing us to measure the contribution from
heavily obscured AGNs over truly cosmological scales.  To achieve this
science goal \nustar\ has undertaken four extragalactic surveys,
spanning a range of different combinations of area and depth, with the
deepest observations identifying more common, faint sources. The
complementary shallower, wider surveys that will cover rarer, more
extreme sources.  The tiers that make up the \nustar\ extragalactic
survey are (a) a large area (currently covering $\approx$7 deg$^2$.),
mostly shallow serendipitous survey consisting of the areas around
targeted sources (described in \citealt{Alexander13} and Lansbury et
al. in prep.), (b) a mid-depth ($\approx90~ks$ maximum unvignetted
depth) survey of the 2 deg$^2$. Cosmic Evolution Survey (COSMOS;
\citealt{Scoville07}), described in \cite{Civano15} and (c) two deep
($\approx360$~ks maximum unvignetted depth), small area surveys of the
$\approx$0.25 deg$^2$. Extended Chandra Deep Field South (hereafter,
ECDFS; \citealt{Lehmer05}, hereafter L05) -- the focus of this study
-- and the $\approx$0.24 deg$^2$.  Extended Groth Strip (i.e., EGS;
\nustar\ analysis to be presented in Aird et al. in prep.).  By
concentrating on extragalactic fields, these surveys will give a
census of hard X-ray sources that is unbiased by pre-selection,
enabling the characterization of a significant sample of the
``typical'' population responsible for the bulk of the CXB.  Indeed,
\nustar\ has already demonstrated this capability in the case of the
ECDFS source NuSTAR J033202-2746.8.\footnote{We note that, due to
  minor changes in our data reduction since the publication of
  \cite{DelMoro14}, the name of this source has been updated to NuSTAR
  J033202-2746.7 in our catalogue.  We note that these changes do not
  affect any of the science results of \cite{DelMoro14}.}  Prior to
\nustar\ the spectral properties of this source were incorrectly
constrained, but it has since been shown to be a high-redshift QSO
with a significant reflection component, despite being Compton-thin
(\citealt{DelMoro14}).  If such reflection is common within the
obscured -- but sub-Compton-thick -- AGN population, it would have a
significant impact on our understanding of the make up of the CXB.

In this study, we describe the \nustar\ observations of the ECDFS that
form one of the two deepest contiguous components of the \nustar\
extragalactic survey (see \S\ref{SS:Obs}).  In \S\ref{SS:Pro} we
describe the data reduction and processing steps we took to form the
final science, background and exposure mosaics. In \S\ref{SS:Sou},
we describe how we obtained our ``blind'' source catalog, the format of
which is described in our Appendix.  In \S\ref{S:Res} we describe the
first results from this sample of sources including derived properties
such as source fluxes, spectral indices and luminosities.  We discuss
constrains on the number of sources not detected by either \chandra\
nor \xmm\ in \S\ref{S:Sam}.  In \S\ref{S:The} we give a brief overview
of the \nustar\ detected sources in the context of the previously
known X-ray sources in this field.  Finally, in \S\ref{S:Dis}, we
summarize our results.  We adopt $H_{0}=71$~km~s$^{-1}$~Mpc$^{-1}$,
$\Omega_{\rm M}=0.27$, and $\Omega_{\Lambda}=0.73$ and use the
AB-magnitude system throughout (where appropriate).

\section{Observations and analyses}
\label{S:Obs}
The \nustar\ ECDFS survey consists of observations from two separate
passes.  Observations making up the first pass were taken between
September and December 2012, while those making up the second pass
were taken roughly six months later between March and April 2013.  The
details of these observations, including aim points, roll angles and
useable exposure times are provided in Table \ref{T:Det}.  In this
section we describe our observing strategy and outline the steps taken
to process and analyze the resulting data.  We note that, in order to
ensure consistency between the different components of the {\it
  NuSTAR} extragalactic surveys, a determined effort was made to
follow the same analysis techniques for both COSMOS
(\citealt{Civano15}) and the ECDFS surveys wherever possible.

\subsection{Observing strategy}
\label{SS:Obs}
\begin{table}
\begin{center}
  \caption{Details of the individual observations that make up the two
    \nustar\ passes of the ECDFS}\label{T:Det}
  \begin{tabular}{@{}lccccc@{}}
\hline
\hline
(1)&(2)&(3)&(4)&(5)&(6)\\
Exp. ID&Obs. Date&R.A.&Dec&Roll Angle&$t_{\rm Exp}$\\
\hline
       1&28 Sep. 2012&52.93&$-27.97$& 85.28&44.9\\
       2&29 Sep. 2012&53.06&$-27.97$& 85.30&45.6\\
       3&30 Sep. 2012&53.18&$-27.97$& 85.30&47.1\\
       4&01 Oct. 2012&53.31&$-27.97$& 85.31&47.0\\
       5&02 Oct. 2012&52.93&$-27.86$& 85.30&46.3\\
       6&04 Oct. 2012&53.06&$-27.86$& 85.31&45.4\\
       7&30 Nov. 2012&53.18&$-27.86$&264.99&47.9\\
       8&01 Dec. 2012&53.31&$-27.86$&264.96&48.0\\
       9&03 Dec. 2012&52.93&$-27.75$&266.96&46.7\\
      10&04 Dec. 2012&53.06&$-27.75$&266.94&47.7\\
      11&05 Dec. 2012&53.18&$-27.75$&266.93&48.0\\
      12&06 Dec. 2012&53.31&$-27.75$&266.88&48.4\\
      13&07 Dec. 2012&52.93&$-27.64$&266.92&48.8\\
      14&08 Dec. 2012&53.06&$-27.64$&266.94&49.2\\
      15&09 Dec. 2012&53.18&$-27.64$&266.94&49.4\\
      16&10 Dec. 2012&53.30&$-27.64$&266.95&46.5\\
      17&15 Mar. 2013&53.30&$-27.64$&351.82&48.6\\
      18&17 Mar. 2013&53.18&$-27.64$&351.81&48.9\\
      19&18 Mar. 2013&53.06&$-27.64$&351.83&48.6\\
      20&19 Mar. 2013&52.93&$-27.64$&351.84&46.1\\
      21&20 Mar. 2013&53.31&$-27.75$&351.83&46.4\\
      22&21 Mar. 2013&53.18&$-27.75$&351.84&46.1\\
      23a&22 Mar. 2013&53.06&$-27.75$&351.86&31.0\\
      23b&23 Mar. 2013&53.06&$-27.75$&351.88&15.2\\
      24&24 Mar. 2013&52.93&$-27.75$&356.88&45.7\\
      25&25 Mar. 2013&53.31&$-27.86$&356.88&46.0\\
      26&26 Mar. 2013&53.18&$-27.86$&356.89&45.9\\
      27&27 Mar. 2013&53.06&$-27.86$&  1.97&45.4\\
      28&28 Mar. 2013&52.93&$-27.86$&  1.95&45.4\\
      29&29 Mar. 2013&53.31&$-27.97$&  1.94&45.3\\
      30&30 Mar. 2013&53.18&$-27.97$&  1.93&45.2\\
      31&31 Mar. 2013&53.06&$-27.97$&  1.93&45.2\\
      32&01 Apr. 2013&52.93&$-27.97$&  1.95&45.0\\
\hline
\end{tabular}

\end{center}
\tablecomments{\\
  (1) Exposure ID. \\
  (2) Start date of the observation. \\
  (3)-(5) The aim point right ascension and declination and roll angle
  of the observation. \\
  (6) The useable exposure time (in ks) of the observation after
  filtering out the flaring events.  Flares only affected Exp. IDs 2, 3, 18 and
  22 (see \S\ref{SSS:Rep}).}
\end{table}

\nustar\ features two independent telescopes with corresponding focal
plane modules (hereafter, referred to as FPMA and FPMB), that
simultaneously observe the same patch of sky during each observation.
Each focal plane has a $\approx$12\arcmin$\times$12\arcmin\ field of
view and consists of four CdZnTe detectors.  The physical pixels are
12\arcsec\ on a side, with the scale subdivided into 2.46\arcsec\
pixels in the software.  While the detectors are sensitive to photon
energies in the range 3-100~keV, the optics of the telescope limit
this range to 3-78.4~keV.\footnote{See \cite{Harrison13} for a
  detailed description of the \nustar\ telescope.} The focusing optics
of \nustar\ give it an unprecedented angular resolution at these hard
\mbox{X-ray} energies, with a tight central ``core'' of
FWHM=18\arcsec\ and a half-power diameter of 58\arcsec, meaning
surveys as deep as the ECDFS are not limited by confusion between
detected sources.  The area covered by the ECDFS (i.e.,
$\approx$30\arcmin$\times$30\arcmin; L05) is larger than the \nustar\
field of view, so we adopted a tiling strategy to provide full
coverage of the field.  A strategy of 16 pointings separated by a
half-detector shift -- forming a 4$\times$4 square -- was chosen based
on the findings of pre-survey tests which suggested that this
optimizes the number of detections in the field.  Since the roll angle
of the observatory is a function of time, the observing schedule was
chosen not only to ensure that the edges of each observation were
roughly aligned, but also to align them with the \chandra\ coverage of
this field.

\begin{figure}
  \plotone{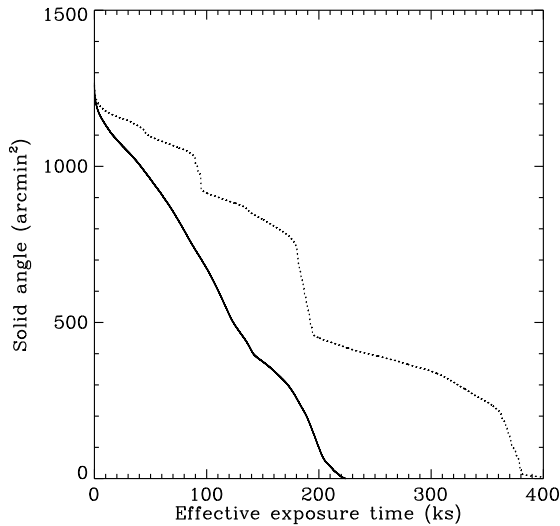}
  \caption{Cumulative area above a given exposure time in the 3-24~keV
    band for the final mosaic in a single focal plane module.  The
    dotted line gives the total area above a given exposure before
    correcting for vignetting, whereas the solid line includes the
    effects of vignetting on the total exposure.}
  \label{F:Cum}
\end{figure}

Each pointing had an effective exposure of approximately 45~\ks\ per
pass, and all but one of the pointings consisted of a single unbroken
exposure, resulting in a total of 33 exposures.\footnote{Exp. ID 23
  was split into two exposures of 31.0 and 15.2~ks each to accommodate
  a Target of Opportunity observation of the Galactic plane source
  NuSTAR J163433-4738.7; see \cite{Tomsick14}} Table \ref{T:Det} gives
the precise useable integration times for each of the 33 separate
exposures after filtering out flaring events (see \S\ref{SSS:Rep}).
The total time spent on the ECDFS across both passes was 1.49~Ms.
On completion of the two passes, the central, deepest
$\approx20$\arcmin$\times20$\arcmin\ region of the field was covered
by eight separate pointings, and thus observed for a total of
$\sim360$~\ks, while the edges had been covered by four pointings
(i.e., to $\sim180$~\ks) and the corners by two (i.e., to
$\sim90$~\ks).  The cumulative area histogram showing the area of the
sky covered to a given exposure time is shown in \fig{F:Cum}.

\subsection{Production of science, background and exposure maps}
\label{SS:Pro}
In this subsection, we outline the detailed steps we took to produce the
final mosaics, identify sources and derive source properties.  All
data reduction was performed using v6.15 of HEASoft, v1.3 of the
\nustar\ developed software, NuSTARDAS (included with v6.15 of
HEASoft), and v4.5 of CIAO.\footnote{HEASoft software available to
  download via http://heasarc.gsfc.nasa.gov/lheasoft; CIAO software
  available to download via http://cxc.harvard.edu/ciao.}

\subsubsection{Reprojection, bad pixels rejection, energy bands.}
\label{SSS:Rep}
Each of the 33 exposures results in two event files, one each for FPMA
and FPMB.  These 66 raw, unfiltered event files were processed and
reprojected using the NuSTARDAS program {\tt nupipeline} to produce 66
clean event files.  Inspection of the cleaned event files revealed
that four exposures (Exp. IDs 2, 3, 18 and 22 in Table \ref{T:Det})
had been significantly affected by flaring events.  These events were
filtered out using CIAO's {\tt dmgti} tool to make a user-defined good
time interval (gti) file, binning in 20~s intervals and rejecting
periods when the average binned count rate exceeded $0.6~{\rm cts\
  s^{-1}}$ (within the whole observable energy range, i.e.,
3-78.4~keV), a threshold selected through visual inspection of the
light curves.  This filtering removed a total of 6.0~ks of exposure
time, representing $\approx0.4\%$ of the total exposure time for the
two passes.  Following \cite{Alexander13} the final cleaned event
files were split into three energy bands, 3-8~keV, 8-24~keV, and
3-24~keV, using HEASoft's {\tt extractor} tool.  The 24~keV upper band
energy is chosen to optimize signal to noise for sources with average
spectral properties.  For these sources, the combination of photon
spectrum falling rapidly with energy, flat background and decreasing
effective area means that detection significance decreases with
increasing high energy cut-off.  As a check, we generated a fourth,
24-40~keV band, but no significant sources were identified using the
detection technique described below.

\subsubsection{Background map generation}
\label{SSS:Bac}

The majority of counts in all of the ECDFS observations are
attributable to background that arises from: (1) an instrumental
background component that dominates at energies $\gtrsim20$~keV; (2) a
focussed background component that is made up of X-ray emitting
sources that are focussed by the telescope optics but lie below the
detection threshold; (3) an ``aperture'' background that is
astronomical in origin, but is due to X-ray photons that have not
scattered off (and thus have not been focussed by) the telescope
optics; (4) a spatially uniform component that is strong at
$\lesssim5$~keV, probably due to neutrons striking the detector; and
(5) another low energy component that is related to solar photons
reflecting off the back of the aperture stop.  The aperture background
dominates the photon counts at $\lesssim 30~{\rm keV}$ and, as such,
most strongly affects our chosen energy bands.

\begin{figure}
  \includegraphics[width=8.5cm]{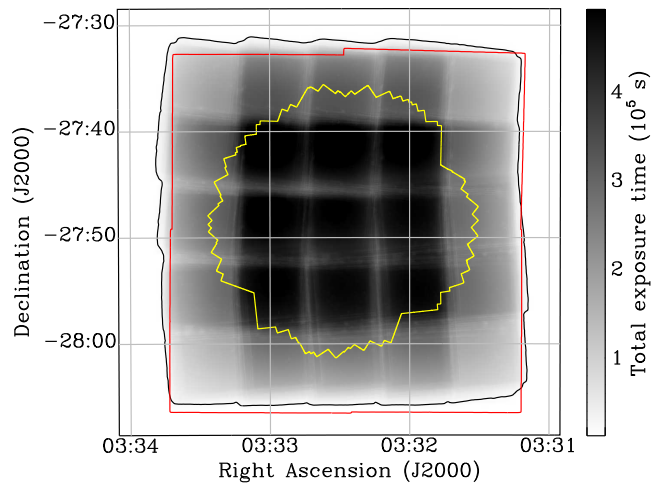}
  \caption{Vignetting-corrected exposure map showing the effective
    exposure of the field, combining both FPMA and FPMB.  Darker grays
    imply longer effective exposures (see colorbar).  The black line
    shows the limit of the \nustar\ exposure, while the red and yellow
    lines show the limits of the 250~ks {\it Chandra} ECDFS (L05) and
    4~Ms CDFS (\citealt{Xue11}) exposures, respectively.}
  \label{F:Vig}
\end{figure}

Due to the asymmetric layout of the optics bench, the aperture
background has a strong gradient that differs between FPMA and FPMB,
making it difficult to subtract by the usual process of extrapolating
from regions around the source.  This is especially true in the case
of overlapping observations such as those that make up the ECDFS
mosaic.  Instead, we use a model of the aperture background for each
FPM produced using the specially developed IDL software {\tt nuskybgd}
(\citealt{Wik14}).  The X-ray spectrum is extracted from a set of
  user-defined background regions and then fit with a pre-defined
  model consisting of the five components listed above (note that
  component $[4]$ is incorporated into the instrumental background,
  i.e., component $[1]$).  A set of pre-determined maps that describe
the spatial distribution of each background component in each energy
band (see \citealt{Wik14}) are then adjusted using the normalizations
of the spectral fit.

When generating the model background, there is the option in {\tt
  nuskybgd} to exclude known bright sources in each image from the
user-defined regions to prevent them from being inappropriately
included in the background estimation.  We experimented with excluding
known bright sources, but they were found to have no significant
impact on either our final detected source list or source
fluxes.\footnote{The list of bright sources to exclude was generated
  using our source detection algorithm with a background created {\it
    without} excluding any sources.} In light of this, and for
reproducibility, we chose not to exclude sources in generating our
final background maps.  Instead, we simply used four large (i.e.,
$\sim5$\arcmin\ on a side) square regions centered on the four chips
that make up each detector to define our background regions (i.e.,
using the same procedure as for the \nustar-COSMOS survey described in
\cite{Civano15}.

A highly accurate synthetic background map is crucial to determine
source reliability and calculate net source counts and, ultimately,
fluxes.  To test the reliability of the synthetic background maps we
calculated the relative difference between the number of counts
extracted from 3,500 30\arcsec\ radius regions distributed
across the ``synthetic'' background and the number of counts extracted
in the same regions distributed across the science mosaics (i.e.,
$(C_{\rm Sci}-C_{\rm Bgd})/C_{\rm Sci}$).  The resulting values are
normally distributed with a dispersion ($\sigma$) of 0.078, 0.064 and
0.053 centered around $1.7\times10^{-3}$, $-1.9\times10^{-3}$ and
$-4.4\times10^{-3}$ for the 3-8~keV, 8-24~keV and 3-24~keV bands,
respectively (FPMA and FPMB combined).  Tests conducted on mock
science images (i.e., Poissonian realizations of the background maps)
demonstrated that this dispersion is consistent with that expected due
to Poisson noise, meaning that the uncertainties introduced by the
background estimation are small compared to those due to the noise in
the data.

\subsubsection{Exposure maps and vignetting}
\label{SSS:Exp}
Exposure maps were generated using the NuSTARDAS software {\tt
  nuexpomap}.  As well as pointing information, this also takes into
account the movement of \nustar 's 10~m long mast when determining the
exposure time of each point on the sky translated into pixel position.
To reduce processing time, there is the option to reduce the number of
calculations by spatially binning the exposure maps.  We therefore
spatially bin our exposure maps over 5 pixels in each dimension, which
is smaller than the 30\arcsec\ aperture used for photometry
measurements and thus has negligible influence on our flux
measurements while speeding up processing by a factor of
$\sim5^{2}=25$.

The effects of vignetting were also taken into account with {\tt
  nuexpomap} to generate effective exposure maps for each observation.
The degree of vignetting is energy-dependent, but generating a
vignetting map for every energy channel is prohibitive.  Instead,
following \cite{Civano15}, we calculated the energy that
minimizes the difference across the three adopted energy bands by
convolving the instrument response with a power-law spectrum of
$\Gamma=1.8$ (i.e., the average of nearby AGN studied at $>10$~keV;
\citealt{Burlon11}).  This results in three energies which we used to
generate the effective exposure maps: 5.42~keV for the 3-8~keV band,
13.02~keV for the 8-24~keV band, and 9.88~keV for the 3-24~keV band.
We calculate that this approximates the vignetting in each energy
channel to within 14.5\% for all three bands, the difference being
largest farthest from the aim point. The effects of applying this
vignetting correction are shown in \Fig{F:Cum}, in which the solid
line shows the cumulative solid angle to a given effective exposure,
evaluated at 3-24~keV.  On average, correcting for vignetting reduces
the nominal exposure time by a factor of $\sim2$.

\subsubsection{Astrometric correction}
\label{SSS:Ast}

\begin{figure*}
  \plotone{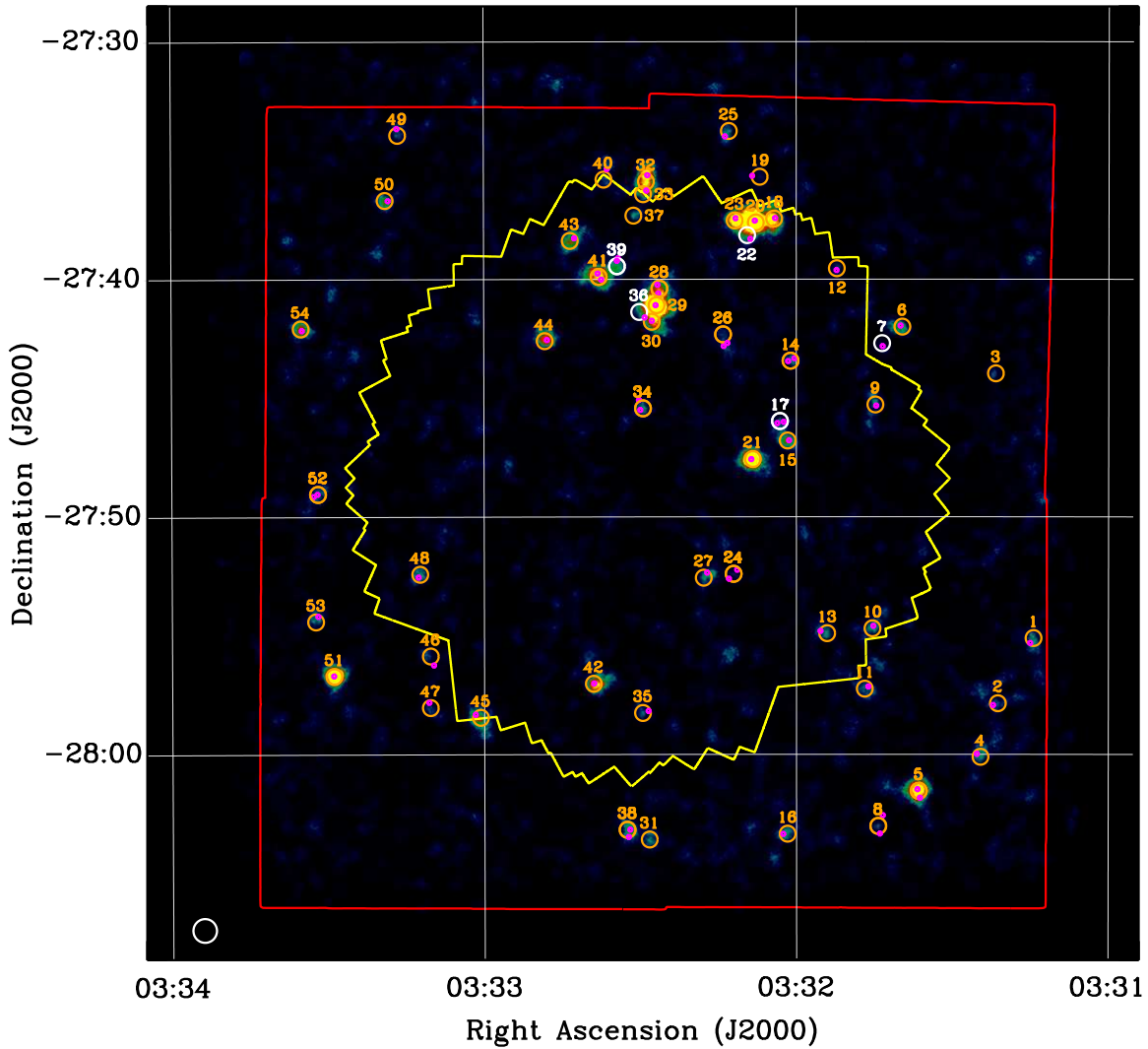}
  \caption{False-probability ($P_{\rm False}$) map produced by
    combining data in the 3-24~keV band from both FPMA and FPMB.
    Blue, green and, finally, yellow shading indicates areas of
    increasingly lower $P_{\rm False}$, and thus the regions most
    likely to be associated with genuine sources.  The yellow and red
    lines indicate the extent of the the 4~Ms CDFS (\citealt{Xue11})
    and 250~ks \chandra-ECDFS (L05) surveys, respectively.  The
      white circle in the bottom-left of the figure indicates the size
      of the 30\arcsec-radius aperture used for measuring source
      photometry.  The positions of all \nustar\ sources found using
    our detection algorithm (see \S\ref{SSS:Bli}) are indicated with
    small numbered circles (the numbers refer to the source ID in the
    online catalog).  Orange circles indicate those that remain
    significant (i.e., meet our significance thresholds) after taking
    into account the effects of neighboring sources (i.e., after
    source deblending; see \S\ref{SSS:Net}), whereas the white circles
    indicate those that no longer prove to be significant.  Note that
    in some instances the position does not correspond to that of the
    minimum $P_{\rm False}$ in the 3-24~keV band.  This is because
    those sources are detected at a higher significance in another
    band and, as such, the position is based on the minimum $P_{\rm
      False}$ position in that other band (see \S\ref{SSS:Bli}).
    Small magenta circles indicate the positions of the \chandra\ or
    \xmm\ counterparts to the \nustar-detected sources.}
  \label{F:Fal}
\end{figure*}

Before combining the individual observations to form the final
science, exposure and background mosaics, we experimented with
correcting for astrometric offsets between individual exposures.  In
brief, we stacked the \nustar\ data at the positions of known {\it
  Chandra}-detected sources and used the relative positions of the
stacked detected source to perform first order (i.e., x-y shift)
astrometric corrections.  However, we found that this correction had
no measurable impact on either the list of detected sources, nor their
measured fluxes.  As such, we chose to not include any astrometric
correction in the rest of our analyses in order to keep the analysis
stream as simple as possible.  Based on their analysis of simulated
\nustar\ observations of the COSMOS field, \cite{Civano15} estimate an
average positional uncertainty of 6.6\arcsec , which, due to the many
similarities between the ECDFS and COSMOS \nustar\ survey strategies
and data reduction, we adopt as the positional uncertainty for the
detected ECDFS sources.

\subsubsection{Mosaicing}
\label{SSS:Mos}
To create a set of contiguous maps, the individual science, background
and exposure images were mosaiced using HEASoft's {\tt ximage}
package.  Our experimentation with astrometry correction (see
\S\ref{SSS:Ast}) demonstrated that the frames from both FPMA and FPMB
are co-aligned to within measurable tolerances, so we are able to
combine the data from these two modules by adding together the
equivalent mosaics, thereby increasing the sensitivity of the final
combined mosaic.  The final mosaiced vignetting-corrected exposure
map, combining FPMA and FPMB, is shown in Figure \ref{F:Vig}.  All
source detection and derived photometric measurements are taken from
the combination of the FPMA and FPMB mosaics.

\begin{figure*}
  \plotone{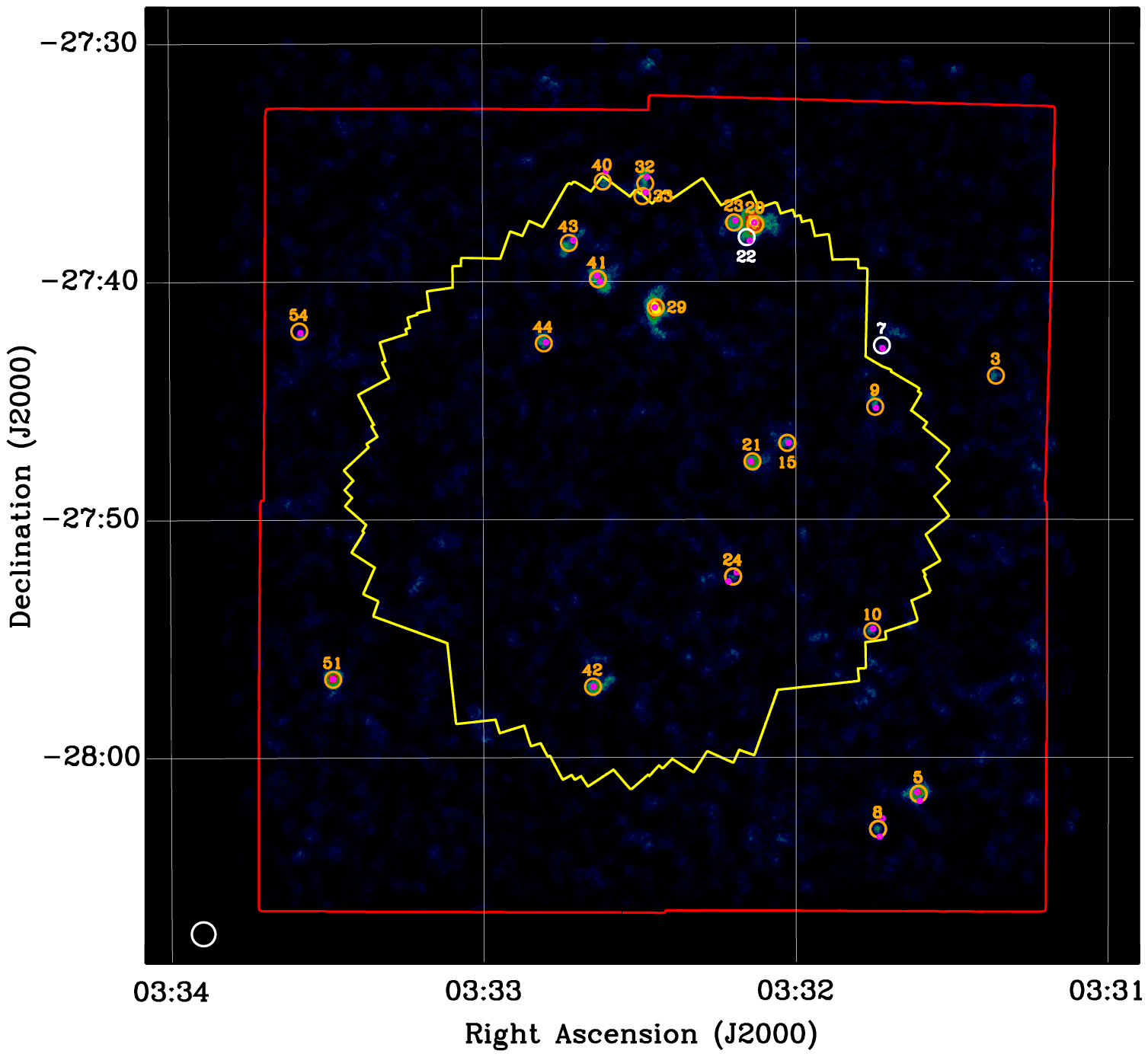}
  \caption{Same as Figure \ref{F:Fal}, but for the 8-24~keV band.
    Sources shown are only those that are detected in this band.}
  \label{F:Fal_HB}
\end{figure*}

\subsection{Source detection and calculation of derived properties}
\label{SS:Sou}
In this subsection we detail how we produced our source catalog from
the final science, background and exposure maps.  We adopted a
multi-stage approach which successfully separates nearby sources and
ensures that all significant sources are identified, while maintaining
a low number of false-positive detections.  The steps are:
\begin{enumerate}
\item we first generate a seed list of potential sources in each band
  using our source detection procedure and a low significance
  detection threshold;
\item we combine the seed lists from each band using positional
  matching and those not meeting our final, more stringent
  \mbox{false-probability} cut in any band are excised from the list
  to form our final catalog;
\item we perform aperture photometry at the positions of the sources
  in our final catalog;
\item we perform deblending (see \S\ref{SSS:Net}) on the sources in
  the final catalog to account for the flux contribution from
  neighboring significant sources; those sources that prove to be
  below our significance threshold post-deblending are flagged, but
  are retained in the final catalog.
\end{enumerate}
A full description of each of these steps is provided below.

\subsubsection{Initial source seed list}
\label{SSS:Bli}
The strong, spatially varying background of the \nustar\ ECDFS mosaics
(see \S\ref{SSS:Bac}) makes blind source detection challenging.
Considerable testing with CIAO's {\tt wavdetect} source-identification
algorithm (\citealt{Freeman02}) revealed that it is not designed to
deal with the deep \nustar\ data in the ECDFS, and it identified large
numbers (i.e., $\sim$200) of false-positive detections, particularly
in the outer regions of the field, likely due to the high background
of \nustar\ images compared to the \chandra\ data for which {\tt
  wavdetect} was originally developed to analyze.  As a consequence,
we instead employed the alternative approach of performing source
detection directly on false-probability (hereafter, $P_{\rm False}$)
maps generated from the mosaics.  These $P_{\rm False}$ maps, which we
calculate using the incomplete Gamma function (see
\citealt{Georgakakis08}), give the likelihood that the signal at each
position in the mosaic is due to random fluctuations in the supplied
background (i.e., not due to a real source).  The maps were produced
by passing smoothed science and background images to IDL's {\tt
  igamma} function, which computes the incomplete Gamma function at
every position in the mosaics, i.e.,
\begin{equation}
  P_{\rm False} = {\tt igamma}({\rm Sci, Bgd})
\end{equation}

\noindent
where Sci and Bgd are the smoothed science and background mosaics,
respectively.  Two different smoothing lengths comprised of 
  10\arcsec\ and 20\arcsec-radius top-hat functions were employed;
the smaller radius helps separate adjacent sources, while the larger
helps to identify fainter sources.  Smoothed $P_{\rm False}$ maps were
generated for each of our three energy bands and source detection was
performed separately on each.  Figure \ref{F:Fal} shows the 20\arcsec
-smoothed $P_{\rm False}$ map for the 3-24~keV band mosaic.

We used the SExtractor source detection algorithm (\citealt{Bertin96})
to identify regions of the $P_{\rm False}$ maps below a given
threshold.  Since SExtractor is designed to identify peaks in an
image, we provided the log of the inverse $P_{\rm False}$ maps as
input.  The SExtractor detection algorithm requires a number of input
parameters, the most important of which for our purposes is the
threshold above which (in the inverse map) a source is considered to
be significant.  Poisson realizations of the blank background maps
(see \S\ref{SSS:Bac}) were used to set the thresholds in each band.
These Poisson realizations were treated as input science images and
analyzed using the same procedures as the real science maps.  By
running our source detection algorithm on 100 realizations of these
simulated, blank maps, we determined the $P_{\rm False}$ value below
which we should expect, on average, one false-positive detection per
ECDFS area per band and set those as our thresholds.  This corresponds
to thresholds of ${\rm log}(P_{\rm False}) \leq -5.19, \leq -5.22$ and
$\leq -5.34$ in the 3-8~keV, 8-24~keV and 3-24~keV bands,
respectively.  We note that these thresholds correspond to
$\approx99\%$ reliability, as evaluated by the simulated \nustar\
observations described in \cite{Civano15}.  Based on our Poisson
realizations, there is a 34\%, 42\%, 19\%, 3\%, 1\% and 1\% likelihood
that the 3-24~keV band mosaic contains 0, 1, 2, 3, 4 and 5 false
positive detections (i.e., have ${\rm log}(P_{\rm False})\leq
-5.34$).\footnote{The corresponding likelihoods for the 3-8~keV band
  are 38\%, 33\%, 24\%, 4\%, 0\% and 1\%, and for the 8-24~keV band
  they are 42\%, 32\%, 18\%, 4\%, 3\% and 1\%.  We note that, in
    all bands, the likelihood-weighted number of false positives sum
    to one, confirming that the average number of false positive
    detections per realisation is one.}  For the generation of our
seed catalog we employ a somewhat more liberal threshold of ${\rm
  log}(P_{\rm False}) \leq -4.5$ in all three bands to ensure that we
detect all potential sources and employ the more stringent thresholds
in our final cut after combining sources detected in each band (see
\S\ref{SSS:Net}).  As the input maps have already been smoothed, we
consider even a single pixel above these thresholds to be a potential
source. The coordinates of each detection are taken as the position of
the local minimum $P_{\rm False}$ (rather than, for example, the
centroid) to maximize the likelihood that it will satisfy our more
stringent final threshold cuts.

As we are performing separate source detection on images smoothed by
two different smoothing lengths, SExtractor often identifies multiple
nearby detections associated with the same source.  If these were left
in our seed catalog, they would lead to over-deblending at the
deblending stage (i.e., the source flux would be erroneously deblended
into these multiple detections).  To account for this, for each seed
detection we identify all other seed detections within a 30\arcsec\
radius (the radius of the aperture used for photometry measurements)
and retain the detection with the lowest $P_{\rm False}$ in our
20\arcsec -smoothed maps (even if initially detected in the
10\arcsec-smoothed maps).  Again, this is done to increase the
likelihood that a detection in our seed catalog will ultimately
satisfy our final significance cut.

We take a similar approach to match sources identified in our three
different energy bands.  For each detection in the 3-24~keV band we
identify the nearest source within a 30\arcsec\ search radius in the
other two energy bands, resulting in up to three different positions
(i.e., one position for each band).  Our final position is taken from
the band giving the lowest $P_{\rm False}$ (based on a 20\arcsec\
radius smoothing) to again maximize the likelihood that it will pass
our final significance cut.  All further analyses (i.e., aperture
photometry, deblending) are performed using this final position.

After combining the seed catalogs from the three separate bands, we
apply our more conservative $P_{\rm False}$ cuts.  For this, we
calculate the seed source $P_{\rm False}$ in each band based on a
20\arcsec\ extraction radius (irrespective of whether the source was
detected at 10\arcsec\ or 20\arcsec-radius smoothing), and remove
those that do not meet our final cuts (i.e., ${\rm log}(P_{\rm False})
\leq -5.19, \leq -5.22$ and $\leq -5.34$ in the 3-8~keV, 8-24~keV and
3-24~keV bands, respectively).  We also remove any sources in areas of
low exposure (i.e., $<4\times10^4$~s, corresponding to $\lesssim10\%$
of the peak vignetted survey exposure).

In total, the above procedure returned 54 sources detected in at least
one band that constitute our final source catalog.  Aperture
photometry was performed at the positions of these significant,
detected sources followed by deblending, as described in the following
subsection.

\subsubsection{Net counts and deblending}
\label{SSS:Net}

We derive net counts, count rates and fluxes at the positions of all
the significantly detected sources identified using the procedure
described above.  To determine total (i.e., source $+$ background)
counts for our detections, we summed the total number of counts within
30\arcsec\ of the final source position in the combined (i.e.,
FPMA$+$FPMB) science mosaics.  This aperture size was chosen as a
compromise between attempting to maintain as low an aperture
correction as possible (a factor of 2.24 for 30\arcsec), and reducing
the level of contamination from neighboring sources.  Furthermore,
tests showed that, compared to using 15\arcsec\ and 45\arcsec\
apertures, 30\arcsec\ led to \nustar\ 3-8~keV fluxes that most closely
matched those derived from {\it Chandra} observations (see
\S\S\ref{SSS:Flu} and \ref{SSS:Mul}).  Total background-only counts
were calculated in the same size apertures centered on the positions
of our detected seeds using the background images described in
\S\ref{SSS:Bac}.  The net number of counts for each detection was
calculated by subtracting the background counts from the total counts.
The upper and lower 1$\sigma$ confidence limits on the total source
counts are calculated following \cite{Gehrels86}.  To this we add the
background error ($\sigma_{\rm Bgd}$) in quadrature with the source
count error to derive the error on the net count rate, scaling
  the background counts by a factor of 125, which is roughly the ratio
  between the area of the 30\arcsec\ aperture used for photometry
  measurements and the total area over which the background model is
  defined (i.e., $=4\times5\arcmin\times5\arcmin$), i.e.,

\begin{equation}
  \sigma_{\rm Bgd}=(1+\sqrt{125C_{\rm Bgd}+0.75})/125
\end{equation}

\noindent
where $C_{\rm Bgd}$ is the total number of background counts extracted
from the background image described in \S\ref{SSS:Bac}.

The relatively extended PSF of \nustar\ means we must factor-in the
contribution from neighboring detections to the measured count rates
and fluxes.  To deblend a given detection in our final catalog we
assume that the net photon counts within the adopted 30\arcsec\ radius
aperture is the sum of that due to that detected source, plus the
contribution from any other \nustar-detected sources within 90\arcsec
(all the sources in the ECDFS are faint enough for any contribution
from sources outside this radius to be considered negligible).  While
we acknowledge that there will also be some contribution from
\nustar-undetected sources, our goal here is to produce a catalog
based solely on \nustar\ data, rather than relying on prior (e.g.,
\chandra\ or \xmm) information.  We make the simplifying assumption
that the contribution of a neighboring source is a function of only
its brightness and separation from the source of interest.  Under this
assumption, the problem of deblending reduces to a set of solvable
linear simultaneous equations; e.g., in the case of three sources:

\begin{eqnarray}
C_{T}^{1} = N(r_{1,1})C_{D}^{1} + N(r_{1,2})C_{D}^{2} +
N(r_{1,3})C_{D}^{3} \nonumber \\
C_{T}^{2} = N(r_{2,1})C_{D}^{1} + N(r_{2,2})C_{D}^{2} +
N(r_{2,3})C_{D}^{3} \\
C_{T}^{3} = N(r_{3,1})C_{D}^{1} + N(r_{3,2})C_{D}^{2} +
N(r_{3,3})C_{D}^{3} \nonumber
\end{eqnarray}

\noindent
where $C_{T}^{n}$ is the total net photon counts within the 30\arcsec\
aperture of source $n$, $C_{D}^{n}$ is the deblended net photon count
of source $n$ (again, within 30\arcsec) and $N(r_{n,m})$ is the
relative normalization that takes into account the separation between
the sources $n$ and $m$ (normalized to 1 at $N(r_{n,n}=0)$).  In
reality, this is complicated by the non-azimuthally symmetric PSF
which lengthens with increasing off-axis angle.  From simulations, we
estimate that the effect of this to the deblended flux of our detected
sources is small compared to photometric uncertainties, i.e.,
typically $\sim8\%$.  Positional uncertainties are not taken into
account in calculating the uncertainties on deblended count rates, but
uncertainties in $C_{T}$ are factored-in using a Monte-Carlo
technique, whereby Gaussian noise (appropriate for the high net counts
of our significantly detected sources) is added to each $C_{T}$
according to the uncertainty on this value.  This is performed
$10^{4}$ times for each source and the resulting distribution (which
is closely approximated with a Gaussian due to regression to the mean)
for $C_D$ is assumed to give the uncertainty on this value.

\begin{table}
\begin{center}
  \caption{Polynomial coefficients for calculating fluxes and $\Gamma$}\label{T:Pol}
  \begin{tabular}{@{}lccccc@{}}
\hline
\hline
(1)&(2)&(3)&(4)&(5)&(6)\\
Property&$a_0$&$a_1$&$a_2$&$a_3$&$a_4$\\
\hline
$f_{\rm SB}$& 6.38& 0.78&-0.32& 0.10&-0.01\\
$f_{\rm HB}$&21.01&-3.12&-0.27& 0.19&-0.02\\
$f_{\rm FB}$&17.13&-4.38&-0.21& 0.42&-0.06\\
$\Gamma$& 1.27&-2.81& 0.04&-0.06&-0.12\\
\hline
\end{tabular}

\end{center}
\tablecomments{\\
  (1) $f_{\rm SB}$, $f_{\rm HB}$, $f_{\rm FB}$ correspond to the
  aperture-corrected fluxes in the
  3-8~keV, 8-24~keV and 3-24~keV bands, respectively, in $10^{-14}~{\rm ergs\ s^{-1}\ cm^{-2}}$.\\
  (2)-(6) The polynomial coefficients used with Eqns. 4 and 5.}
\end{table}

At this stage we also perform deblending assuming a 20\arcsec-radius
aperture (compared to the 30\arcsec\ used above for aperture
photometry); recall that initial source detection is performed using
both 10\arcsec\ and 20\arcsec-radius smoothing (see
\S\ref{SSS:Bli}). We then use these 20\arcsec\ deblended counts, in
conjunction with the background photon counts, to re-calculate the
$P_{\rm False}$ of each source post-deblending and flag (but not
remove) those that are no longer significant after deblending.  Of the
54 sources in the final catalog, five are flagged as being no longer
significant post-deblending, leaving 49 sources that are significant
post-deblending.

From the deblended 30\arcsec -aperture net photon counts we calculate
net count rates (and associated uncertainties) using the mean combined
(i.e., the sum of both detectors) effective exposure time within a
30\arcsec\ aperture of the detected seed position. Deblended 8-24~keV
to 3-8~keV band ratios are calculated using the Bayesian Estimation of
Hardness Ratios (BEHR) method described in
\cite{Park06}.\footnote{BEHR code available from:
  http://hea-www.harvard.edu/astrostat/behr/} We report the median and
upper and lower 68th percentiles (i.e., $1\sigma$ uncertainties)
returned by this method. Photon indices and fluxes were calculated
from these deblended band ratios and net count rates following the
procedure described in the next subsection.

\subsubsection{Flux calculation}
\label{SSS:Flu}
Following the same basic approach as \cite{Alexander13}, observed
deblended fluxes in each band and effective photon indices (i.e.,
$\Gamma$) were calculated from the deblended 30\arcsec\ aperture count
rates using conversion factors derived from XSPEC model spectra.  To
generate the conversion factors, we use XSPEC's {\tt fakeit} command
to model power-law spectra spanning a range of power-law indices
($0.0<\Gamma<3.0$, in increments of 0.01; XSPEC model: {\tt pow}) and
taking the rmf and arf at the aim-point of the two detectors into
account.  ``Fake'' fluxes and count rates in the three bands were
extracted from these synthetic spectra and were used to generate
polynomial solutions that relate observed fluxes and photon indices to
our observed count rates and 8-24~keV to 3-8~keV band ratios:

\begin{figure}
  \plotone{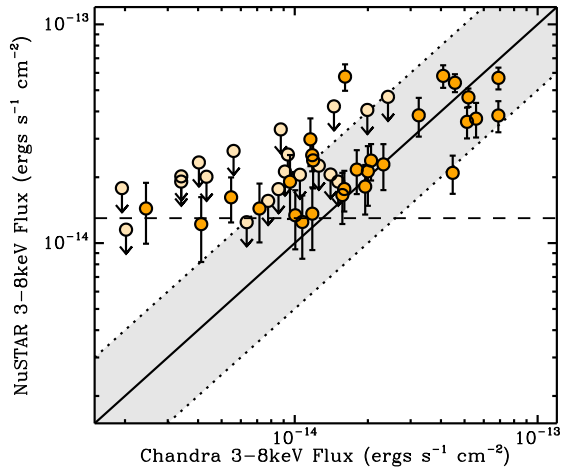}
  \caption{Comparison of deblended, aperture-corrected \nustar\
    3-8~keV fluxes versus the \chandra-measured fluxes of all the
    \chandra\ sources within 30\arcsec\ of the \nustar\ source
    position. Filled circles represent 3$\sigma$ detections, whereas
    open circles represent 3$\sigma$ upper limits.  The solid line
    shows the 1:1 relation, while the grey shading indicates the
    region within a factor of 3 of this line.  The dashed horizontal
    line indicates the sensitivity limit of the deepest region of the
    survey in the 3-8~keV band, i.e., $\approx1.3\times10^{-14}~{\rm
      ergs~s^{-1}~cm^{-2}}$.}
  \label{F:Comp}
\end{figure}

 \begin{equation}
   \frac{f_{\rm E}}{R_{\rm E}}=10^{-14}\sum\limits_{i=0}^n a_i\Gamma^i 
\label{fpoly} 
\end{equation}
\noindent
where
\begin{equation}
  \Gamma=\sum\limits_{i=0}^n a_i b^i 
\label{gpoly} 
\end{equation}

\noindent
and $f_{\rm E}$ is the flux within a given energy band (in ${\rm ergs\
  s^{-1}\ cm^{-2}}$), $R_{\rm E}$ is the count rate in the same energy
band (in ${\rm ks^{-1}}$), $b={\rm log}\left(R_{\rm 8-24\ keV}/R_{\rm
    3-8\ keV}\right)$ (i.e., the logarithm of the 8-24~keV to 3-8~keV
band ratio) and $\Gamma$ is the observed (i.e., not corrected for
absorption) photon index.  The polynomial coefficients for the above
equations, $a_i$, are given in Table~\ref{T:Pol} and reproduce the
XSPEC fluxes and photon indices to within $1\%$ across $-0.7<b<0.4$
(corresponding to $0.0<\Gamma<3.0$).  Where a source is detected in
both the 3-8~keV and 8-24~keV bands (and thus has a well constrained
photon index), we use the derived photon index, otherwise we assume a
fixed $\Gamma$ of 1.8 (corresponding to count rate to flux conversions
of $7.2\times10^{-14}$, $1.5\times10^{-13}$ and $1.0\times10^{-13}$
${\rm ergs\ s^{-1}\ cm^{-2}\ ks}$ for the 3-8~keV, 8-24~keV and
3-24~keV bands, respectively).  These flux conversions take the size
of the aperture into account to return aperture-corrected fluxes
(which are reported throughout).

To verify the accuracy of our process for calculating deblended
\nustar\ fluxes, we carry out a comparison between the \nustar-derived
3-8~keV fluxes and the total \chandra-derived 3-8~keV fluxes arising
from all \chandra\ sources within 30\arcsec\ of the \nustar\ source
position combined.  This comparison is shown in \fig{F:Comp}.  The
\nustar\ fluxes cluster along the 1:1 line shown on this plot and all
but two \nustar\ sources are consistent (within $3\sigma$ errors) with
the total \chandra\ flux to within a factor of two, thereby validating
our approach of calculating \nustar\ fluxes.  We note that some of the
scatter in this plot is expected to be introduced by intrinsic source
variability (e.g., \citealt{Paolillo04, Young12}) and spurious matches
between the \nustar\ and \chandra\ sources (see \S\ref{SSS:Mul}).

\begin{figure}
  \plotone{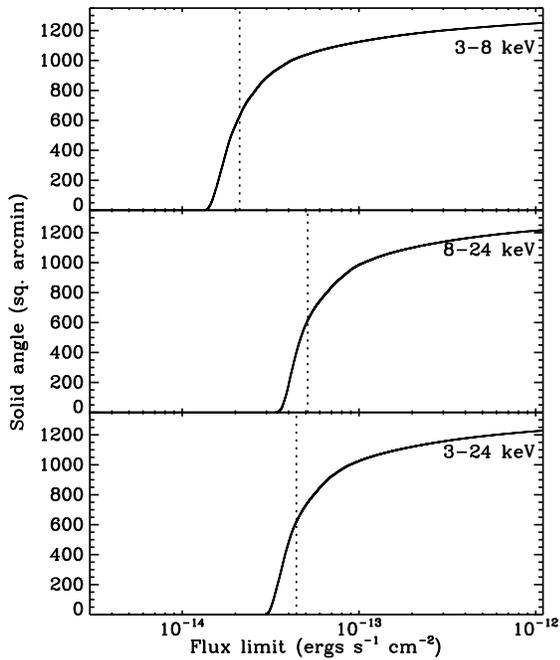}
  \caption{Plot showing the solid angle of the survey below a given
    aperture-corrected flux sensitivity (shown on abscissa)
    corresponding to a S/N of 3 in each of our \nustar\ bands.  These
    sensitivity limits were calculated from the background and
    exposure mosaics alone using the procedure outlined in
    \S\ref{SSS:Flu}.  The most sensitive region in the 3-8~keV,
    8-24~keV and 3-24~keV bands reach $\approx1.3\times10^{-14}$,
    $\approx3.4\times10^{-14}$ and $\approx3.0\times10^{-14}~{\rm
      ergs~s^{-1}~cm^{-2}}$.  The dotted lines indicate the flux
    limits for an area of $\approx630$ arcmin$^2$, which is roughly
    half the solid angle of the full \nustar\ ECDFS survey, and
    corresponds to values of $\approx2.1\times10^{-14}$,
    $\approx5.1\times10^{-14}$ and $\approx4.4\times10^{-14}~{\rm
      ergs~s^{-1}~cm^{-2}}$ for the 3-8~keV, 8-24~keV and 3-24~keV
    bands, respectively.}
  \label{F:Plo}
\end{figure}

We can calculate the limiting flux at every point in the final mosaics
using the background and exposure maps assuming the detection $P_{\rm
  False}$ thresholds introduced in \S\ref{SSS:Bli} (i.e., ${\rm
  log}(P_{\rm False}) \leq -5.19, \leq -5.22$ and $\leq -5.34$ in the
3-8~keV, 8-24~keV and 3-24~keV bands, respectively) and using the
count rate to flux conversions above, assuming $\Gamma=1.8$.  We use
this to determine the area of the survey reaching a given flux limit
in each of our bands, the results of which are shown in \fig{F:Plo}.
The most sensitive regions of the survey reach
$\approx1.3\times10^{-14}$, $\approx3.4\times10^{-14}$ and
$\approx3.0\times10^{-14}~{\rm ergs~s^{-1}~cm^{-2}}$ in the 3-8~keV,
8-24~keV and 3-24~keV bands, respectively.  We note, however, that
these are the theoretical flux limits of the survey for isolated
sources.  In reality, the relatively large \nustar\ PSF may affect the
total area to a given limit as regions around bright sources are less
sensitive due to contamination and issues arising from deblending.  A
full understanding of these second-order factors will be necessary to
obtain, e.g., source number counts and luminosity functions, to be
published in Harrison et al. (in prep.) and Aird et al. (in prep.).

\subsubsection{Multiwavelength counterparts, redshifts and luminosity
  determination}
\label{SSS:Mul}
A major benefit of observing the ECDFS with \nustar\ is the wealth of
ancillary multiwavelength data available for the field, making the
characterization of identified sources comparatively straightforward.
As much of these ancillary data have already been matched to sources
identified in the \chandra\ and XMM-{\it Newton} observations of this
field (i.e., the \chandra\ 250~ks ECDFS, \chandra\ 4~Ms CDFS and XMM-{\it
  Newton} 3~Ms ECDFS surveys; described in L05, \citealt{Xue11} and
\citealt{Ranalli13}, respectively), we obtain this information by
matching to these previous X-ray catalogs.  We first match to the
\chandra -250~ks ECDFS catalog (L05) using a 30\arcsec\
search radius.  We report all matches that contribute at least 20\%\
of the total flux from all \chandra\ sources within 30\arcsec\ of the
\nustar\ position (i.e., we ignore those faint \chandra\ sources that
do not contribute significantly to the total \chandra\ flux within the
search radius).  Following this prescription, of the 54 sources in our
final catalog, 48 were found to have at least one \chandra\
counterpart within the search radius.  Of these, twelve \nustar\
sources were found to have two \chandra\ matches. No \nustar\ source
was found to have more than two \chandra\ matches within the search
radius.  Of the 49 \nustar\ sources that are significant
post-deblending, 44 have at least one \chandra\ counterpart.

Considering all 809 \chandra\ sources listed in L05 with
our 30\arcsec\ matching radius corresponds to a high spurious matching
fraction of $\sim70\%$ for our sample.  However, we note that the
majority (i.e., $\sim85\%$ of matched \chandra\ counterparts have
\chandra\ 3-8~keV fluxes $>2\times10^{-15}~{\rm ergs~s^{-1}~cm^{-2}}$,
of which there are 289 in the full L05 catalogue.
Considering only these sources corresponds to a spurious matching
fraction of $\sim25\%$, which we consider to be a more reasonable
estimation for the spurious matching fraction for our sample.  Of
course, following this logic, brighter \chandra\ counterparts are less
likely to be spurious than fainter sources; we estimate the spurious
matching fraction of $F_{\rm 3-8keV}>10^{-14}~{\rm
  ergs~s^{-1}~cm^{-2}}$ \chandra\ sources to be $\sim3\%$, compared to
$\sim40\%$ for $F_{\rm 3-8keV}>10^{-15}~{\rm ergs~s^{-1}~cm^{-2}}$.
    
The six (of 54) sources without \chandra\ 250~ks counterparts were
then matched against the \chandra -4~Ms CDFS catalog (again, using a
30\arcsec\ search radius), which led to two further matches (and no
multiple matches).  Finally, the remaining four \nustar\ sources
without counterparts were matched against the XMM-{\it Newton} 3~Ms
catalog, which resulted in one further match.  As a result, only three
\nustar\ sources have no \chandra\ or XMM-{\it Newton} counterpart,
all three of which are significant post-deblending.

Where a \chandra\ or XMM-{\it Newton} counterpart is identified, the
associated optical counterpart is taken from the respective catalog
(i.e., L05, \citealt{Xue11} or \citealt{Ranalli13} for
\chandra -ECDFS, CDFS and XMM-{\it Newton}-ECDFS, respectively),
together with its associated redshift, if available.  We adopt the
spectroscopic redshift in preference to the photometric redshift in
cases where both are available. Of the 51 sources for which there are
either \chandra\ or XMM-{\it Newton} counterparts, 46 have
spectroscopic redshifts and three have only photometric redshifts
(i.e., giving 49 in total).  The corresponding numbers for those 49
sources that are significant post-deblending are: 41 with
spectroscopic redshifts and three with photometric redshifts (i.e., 44
in total).

For those sources for which either spectroscopic or photometric
redshifts are available, rest-frame $10-40$~keV (non-absorption
corrected) luminosities were calculated from the 8-24~keV
fluxes. $k$-corrections were performed by adopting the derived photon
index for sources that significantly detected in both the 3-8 and
8-24~keV band, otherwise $\Gamma=1.8$ is assumed.  In this work, we do
not attempt to correct the luminosities for the effects of absorption;
this will be the focus of a later study to combine \chandra\ and/or
XMM-{\it Newton} data with the \nustar\ data to obtain the most
reliable absorbing column densities (and hence, corrected
luminosities) currently achievable (Del Moro et al. in prep;
Zappacosta et al. in prep.).

\section{Results}
\label{S:Res}

\begin{figure}
  \plotone{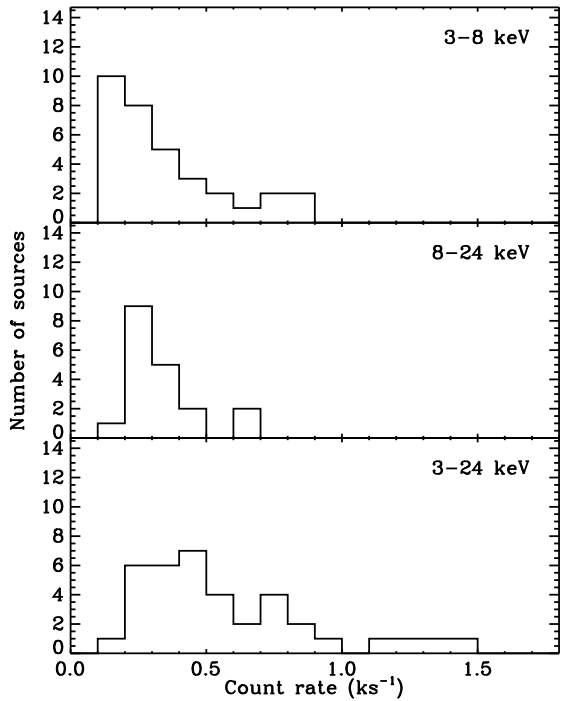}
  \caption{Distributions of non-aperture corrected count rates in our
    three bands for the sources in the ``blind'' catalog.  Here, we
    only include the 49 sources that pass our significance cuts
    post-deblending}
  \label{F:Dis}
\end{figure}

In this section, we describe the properties of the detected sources
and compare them against local (i.e., $z<0.3$) hard X-ray sources
detected in the \swift-BAT survey and those in the \nustar\
serendipitous survey (\citealt{Alexander13}).  We also highlight three
\nustar\ sources that are undetected in both the \chandra\ and \xmm\
coverage of the ECDFS and CDFS, and which may therefore represent
previously unknown contributors to the hard X-ray background.

\subsection{Basic properties}
\label{SS:Bas}
As reported in \S\ref{SSS:Bli}, our final catalog contains 54 sources
that are initially detected as significant in at least one of the
three standard bands.  However, of these 54, five are not significant
after the effects of neighboring sources have been taken into account
(i.e., after deblending).  While these five are retained in the
electronic catalog, the results described in the remainder of this
paper consider only the 49 sources that are significant
post-deblending.

It should also be noted that, since source detection is separate from
photometry, some of our significantly detected sources (i.e., nine)
have $<3\sigma$ counts in all three bands. As they pass our formal
$P_{\rm False}$ threshold, these sources are retained in the
electronic catalog and are considered in our general analyses and
histograms, but are shown as $3\sigma$ upper limits in Cartesian
plots.

Of the 49 sources that make up our final, post-deblended catalog, 12
are significant (i.e., satisfy the $P_{\rm False}$ cuts outlined in
\S\ref{SSS:Bli}) in all three bands, 16 in exactly two (13 in
3-8~keV$+$3-24~keV, 3 in 8-24~keV$+$3-24~keV) and 8, 4, and 9 in the
3-8~keV, 8-24~keV and 3-24~keV bands only, respectively.  As such, 19
are detected in the $8-24$~keV band, i.e., the photon energy range
probed to unique depths by \nustar.  We compare these numbers of
detected sources to those predicted by the X-ray background synthesis
model described in \cite{Ballantyne11}, updated to the \cite{Ueda14}
luminosity function and using an AGN spectrum closely following that
described in \cite{Ballantyne14}, which assumes
$\langle\Gamma\rangle=1.85$ and a \cite{Burlon11} $N_{\rm H}$
distribution.  Convolving this model with the \nustar\ ECDFS
sensitivity curve predicts $\approx$25, $\approx$13 and $\approx$28
sources in the 3-8~keV, 8-24~keV and 3-24~keV bands, respectively.
With 33, 19 and 37 detected sources in the 3-8~keV, 8-24~keV and
3-24~keV bands, respectively, this model under-predicts the actual
number of detected sources by a modest amount, i.e., $\approx$25-30\%.
Interestingly, the biggest percentage difference between the number of
predicted and detected sources is at 8-24~keV, possibly suggesting a
deviation from the adopted model AGN spectrum at these newly probed
energies, although we note that this comparison will be affected by
the small numbers of our sample and field-to-field variations.

\begin{figure}
  \plotone{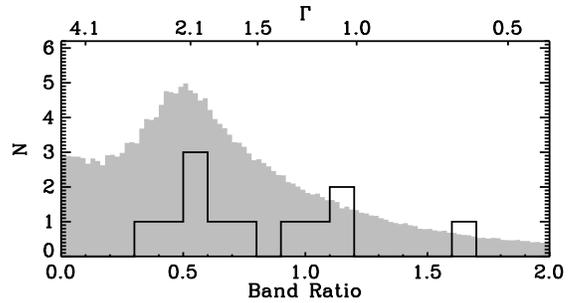}
  \caption{8-24\ keV to 3-8\ keV band ratio (bottom axis) and $\Gamma$
    (top axis) distributions for our sample of ``blind'' sources.  The
    solid lines indicate the distribution of sources that pass our
    significance threshold in both bands after deblending.  The grey
    shaded histograms show the distribution resulting from a Monte
    Carlo analyses of all 49 sources that pass our significance cut in
    at least one band after deblending, incorporating the
    uncertainties on the individual band ratios and photon indices
    (see \S\ref{SS:Bas}).}
  \label{F:Har}
\end{figure}

The distributions of source count rates in each of our bands are
plotted in Figure \ref{F:Dis}. Here, we include all 49 significant
sources, irrespective of their S/N based on our 30\arcsec\ photometry
measurement.  The number of significant sources peaks at $\sim0.2~{\rm
  ks^{-1}}$, $\sim0.1~{\rm ks^{-1}}$ and $\sim0.3~{\rm ks^{-1}}$ in
the 3-8~keV, 8-24~keV and 3-24~keV bands, respectively.  The lowest
number of deblended net source counts in each band are 38, 37 and 36
in the 3-8~keV, 8-24~keV and 3-24~keV band, respectively, and
correspond to two separate sources (NuSTAR J033121-2757.8 [source ID 2
in our table] has the lowest number of counts in the 3-8~keV band,
whereas NuSTAR J033144-2803.0 [ID: 8] has the lowest number of counts
in both the 8-24~keV and 3-24~keV bands etc.).  Conversely, the
highest net source counts in each band are 365, 197 and 545,
respectively, and do correspond to the same source (NuSTAR
J530642-2741.0 [ID: 29]).

Figure \ref{F:Har} shows the distribution of 8-24~keV to 3-8~keV band
ratios for all sources that pass our significance threshold
(post-deblending) in both the 8-24~keV and 3-8~keV bands.  Shown on
the top axis of this plot is the corresponding $\Gamma$ value.  As the
range of band ratios is sparsely sampled, it is difficult to interpret
from the significant sources alone where the band ratio distribution
of our detected sources peaks.  Further insight into the distribution
of band ratios for all our detected sources can be gained by taking a
Monte Carlo approach to account for uncertainties in their count
rates.  We generate $10^4$ mock band ratios for each detection by
randomly sampling the band ratio probability density profiles output
by the BEHR code.  The resulting band ratio distribution is shown in
grey in \fig{F:Har} behind the histogram of those sources
significantly detected in both bands and peaks at $\sim0.5$,
corresponding to $\Gamma\sim2.1$.  We stress that this distribution
only samples the detected sources and may not be representative of the
underlying band ratio distribution of the entire AGN population, which
may be probed using more advanced techniques that are beyond the scope
of this work (e.g., stacking on the positions of known AGN populations
from, e.g., \chandra\ observations).

\begin{figure}
  \plotone{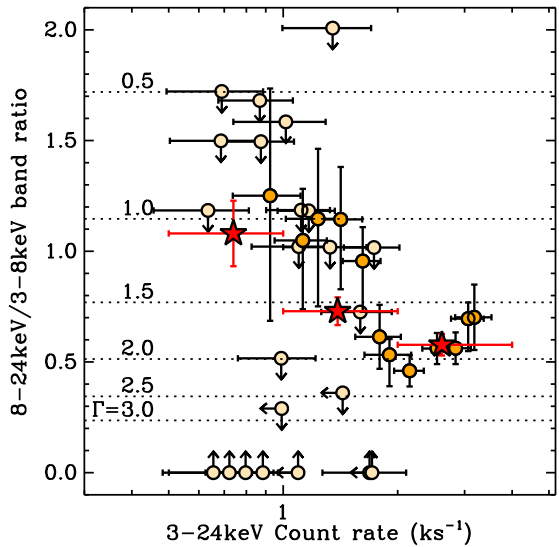}
  \caption{8-24~keV to 3-8~keV band ratio plotted against the
    aperture-corrected 3-24~keV band count rates of our detected
    sources.  The average band ratios in three separate bins of
    3-24~keV band count rate, calculated by summing the counts from
    {\it all} sources in each bin to obtain a stacked detection in
    each band, are shown as red stars.  Horizontal dotted lines
    indicate the photon index at various band ratios.}
  \label{F:8-24}
\end{figure}

Previous studies of deep X-ray surveys have reported an
anti-correlation between the band ratio and count rate in the lower
energy bands (i.e., 2-8~keV to 0.5-2~keV vs. 0.5-8~keV count rates)
for AGN-dominated sources (e.g., \citealt{dellaCeca99, Ueda99,
  Mushotzky00, Tozzi01, Alexander03}).  This trend has been attributed
to an increase in the number of absorbed AGNs detected at lower count
rates.  To investigate whether we see such a trend in the \nustar\
data we plot the 8-24~keV to 3-8~keV band ratio as a function of
3-24~keV count rate in \fig{F:8-24}.  The 12 sources that are detected
at $>3\sigma$ in both the 8-24~keV and 3-8~keV bands appear to show a
weak trend toward higher band ratios at low 3-24~keV count rates.
However, the large number of sources with upper limits in either band
and the narrow dynamical range in 3-24~keV count rates of those
sources detected in all three bands makes it difficult to assess the
significance of this trend.  To investigate this possible trend, we
determine average count rates in three 3-24~keV count rate bins, i.e.,
$0.5\leq R_{\rm FB}/{\rm ks^{-1}}<1$, $1\leq R_{\rm FB}/{\rm
  ks^{-1}}<2$, $2\leq R_{\rm FB}/{\rm ks^{-1}}<4$, getting $>3\sigma$
average count rates for all bands in each bin.  The results of this
averaging is shown in Figure \ref{F:8-24} and shows some evidence of a
decreasing hardness ratio with increasing 3-24~keV count rate,
supporting previous claims from studies of lower energy X-ray bands.
However, we stress that this new result is only based on three count
rate bins and confirmation will be needed by extending the dynamical
range in average count rates via, e.g., stacking the \nustar\
undetected population, which will be the focus of a future study
(e.g., Hickox et al. in prep.).
 
\begin{figure}
  \plotone{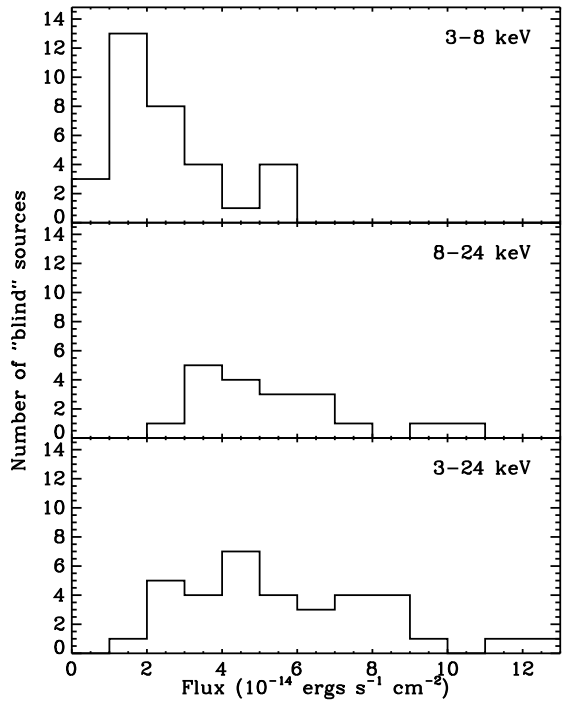}
  \caption{Distributions of aperture-corrected fluxes in our three
    bands for the sources in the ``blind'' catalog.  Here, we only
    include the 49 sources that pass our significance threshold
    post-deblending.}
  \label{F:Dist}
\end{figure}

In the electronic catalog, we also provide derived properties for each
of our detected sources, namely the photon index, $\Gamma$, and the
observed flux in each band (i.e., not corrected for absorption).  The
twelve sources that pass our significance thresholds in both the
8-24~keV and 3-8~keV bands cover a wide range of observed (i.e.,
uncorrected) photon indices, ranging from $0.51^{+0.89}_{-0.24}$ to
$2.27^{+0.47}_{-0.37}$ ($1\sigma$ errors; see Figure \ref{F:Har}). The
majority (i.e., 7 of 12) of sources in our sample for which we can
constrain a photon index have $\Gamma<1.8$ (i.e., the typically
assumed intrinsic AGN photon index), suggesting that a large fraction
of the sources have significant obscuration causing the observed
photon index to harden.  The median photon index and $\pm1\sigma$
interval for these twelve sources is $\overline{\Gamma}=1.70\pm0.52$.
We use the same Monte Carlo simulations as described earlier in this
section (i.e., those used to incorporate uncertainties in the band
ratio distribution) to calculate the median photon index all sources,
including those not individually detected in both bands, finding a
somewhat softer $\overline{\Gamma}=1.90\pm0.53$.

\begin{figure}
  \plotone{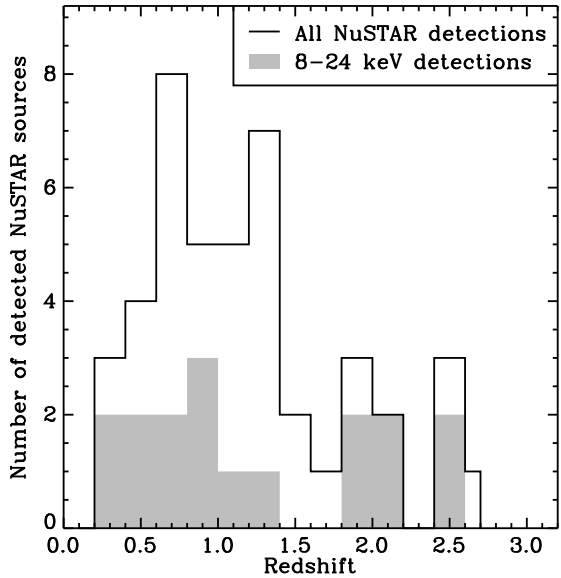}
  \caption{Redshift distribution of the 44 \nustar\ sources that are
    significantly detected (post-deblending) in at least one band and
    have either an associated spectroscopic or photometric redshift
    (solid line).  The shaded histogram shows the same for the 17
    sources significantly detected (post-deblending) in the 8-24~keV
    band that have either a spectroscopic or photometric redshift.}
  \label{F:Reds}
\end{figure}

\begin{figure}
  \plotone{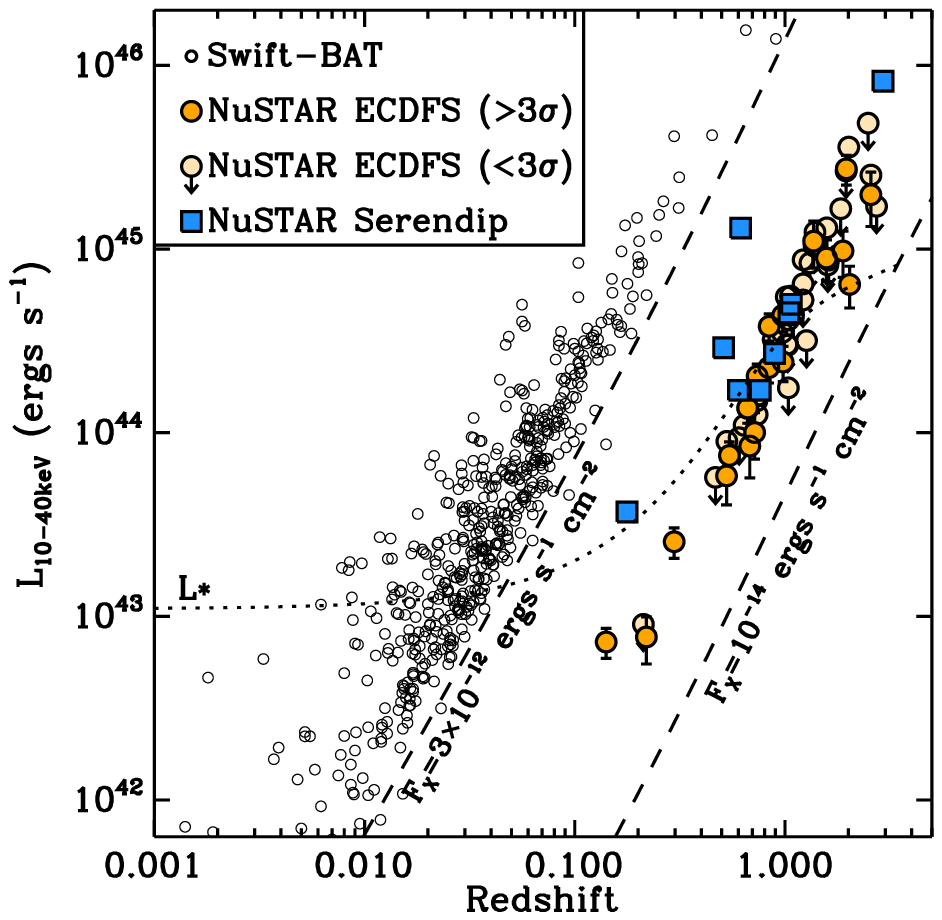}
  \caption{X-ray luminosity in the rest-frame 10-40~keV band
      versus redshift for several hard X-ray selected samples.  We
    show the NuSTAR-ECDFS sources (i.e., our sample), the first ten
    sources from the \nustar\ serendipitous survey
    (\citealt{Alexander13}) and the sources from the \swift-BAT survey
    (\citealt{Baumgartner13}).  In the case of the NuSTAR-ECDFS
      sources, the 10-40~keV luminosity is derived from the observed
      8-24~keV flux, assuming either the measured photon index
      ($\Gamma$) if the source is significantly detected in both the
      3-8~keV and 8-24~keV bands or $\Gamma=1.8$ otherwise.  Upper
      limits are shown for NuSTAR-ECDFS sources when the measured flux
      in the 8-24~keV band is $<3\sigma$.  The dotted line indicates
    the evolution of the position of the knee of the X-ray luminosity
    function, extrapolated from the 2-10~keV luminosity functions of
    \cite{Aird10} assuming $\Gamma=1.8$.  The \nustar\ sources in the
    ECDFS probe to this ``knee'' out to $z\sim1$.}
  \label{F:Lx-z}
\end{figure}

In Figure \ref{F:Dist} we plot the distribution of fluxes of our
detected sources in each of the three bands.  The faintest sources in
the 3-8~keV, 8-24~keV and 3-24~keV bands have fluxes of
$(8.3\pm3.6)\times10^{-15}$, $(2.2\pm1.3)\times10^{-14}$ and
$(1.59\pm0.84)\times10^{-14}~{\rm ergs~s^{-1}~cm^{-2}}$, respectively,
giving an indication of the approximate ultimate sensitivity limit of
the survey in these bands.  However, while passing our
  significance thresholds, these sources all have flux measurements
  below $3\sigma$.  The faintest sources with $>3\sigma$ flux
  measurements have fluxes of $(1.22\pm0.40)\times10^{-15}$,
  $(3.4\pm1.1)\times10^{-14}$ and $(2.36\pm0.74)\times10^{-14}~{\rm
    ergs~s^{-1}~cm^{-2}}$ in the 3-8~keV, 8-24~keV and 3-24~keV bands,
  respectively.  The 8-24~keV and 3-24~keV band flux limits are
roughly 2 orders of magnitude fainter than those of the most sensitive
observations of the \swift-BAT all-sky survey
(\citealt{Baumgartner13}), the previous deepest hard X-ray survey
prior to the launch of \nustar.  We note that the flux limit of the
ECDFS survey is comparable to that of the six deepest observations
that make up the \nustar\ Serendipitous survey
(\citealt{Alexander13}), although the ECDFS survey reaches this depth
over a larger contiguous area and benefits from more comprehensive
ancillary multiwavelength data, an aspect that is exploited in the
next subsection.

The 44 sources in our sample with known redshifts span the redshift
range $z=0.22-2.7$.  The 10-40~keV luminosity range of the 19 of these
44 that are significantly detected in the 8-24~keV band (from
which we calculated 10-40~keV luminosities) span the range
$\sim8\times10^{42}~{\rm ergs~s^{-1}}$ to $\sim3\times10^{45}~{\rm
  ergs~s^{-1}}$.  The redshift-luminosity plane for our sample,
together with comparison samples from the \swift-BAT survey
(\citealt{Baumgartner13}) and the \nustar\ serendipitous survey
(\citealt{Alexander13}) is shown in \fig{F:Lx-z} and highlights the
complementary regions of parameter space that these various surveys
cover, which is the focus of the following subsection.

\subsection{Comparison with low-z samples}
\label{SS:Com}

\begin{figure*}
  \plotone{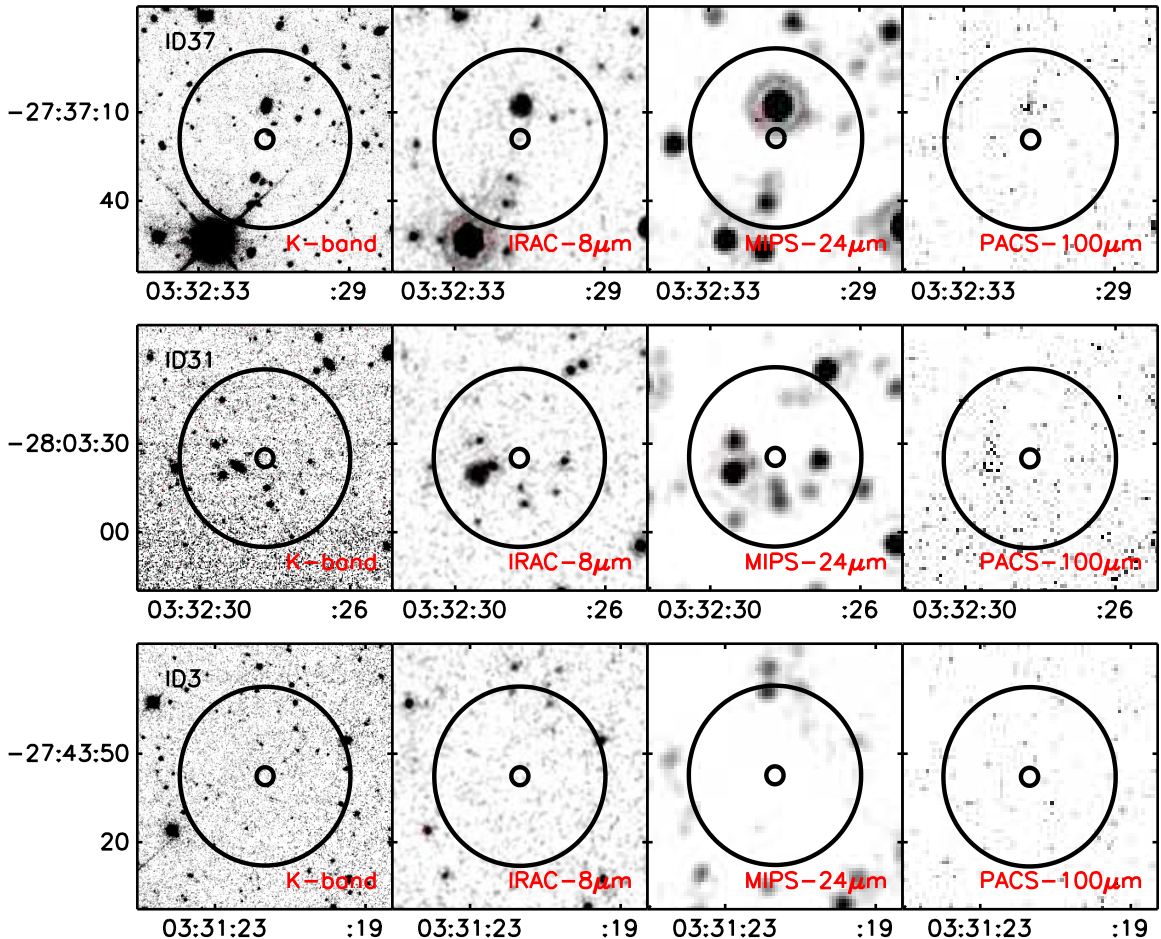}
  \caption{Thumbnail images of the 90\arcsec$\times$90\arcsec\ patches
    of sky at various wavelengths centered on the three \nustar\
    sources without \chandra\ or \xmm\ counterparts.  The smaller
    3\arcsec -radius circle is used to indicate the location of the
    \nustar\ source, while the larger circle shows the 30\arcsec\
    search radius we use when identifying counterparts at other
    wavelengths.  The data used to create these images come from TENIS
    ({\it K}-band; \citealt{Hsieh12}), SIMPLE ({\it Spitzer}-IRAC 
      8~\mum\ $[$ i.e., Ch. 4$]$; \citealt{Damen11}), FIDEL ({\it
      Spitzer}-MIPS 24~\mum) and PEP ({\it Herschel}-PACS;
    \citealt{Lutz11}) surveys.  Only one of these, \nustar\ -ID: 37
    shows any evidence of hosting an AGN at any wavelength other than
    hard X-rays, which comes from mid to far-infrared SED fitting (see
    Figure \ref{F:SED}).}
  \label{F:Thumbs}
\end{figure*}

The AGNs detected and characterized as part of the all-sky \swift-BAT
survey form the local sample that is most directly comparable to the
\nustar-ECDFS AGNs reported here.  \Fig{F:Lx-z} shows that, while the
range in luminosities of both samples are large (i.e., each spanning
roughly 3 orders of magnitude or more) and there is some overlap, the
AGNs in the ECDFS survey are typically an order of magnitude more
luminous than those in the \swift-BAT sample.  As expected, the
difference in the redshift distributions of the two samples is also
large, with the ECDFS sample centered around $z\sim1$, compared to
$z\sim0.03$ for the \swift-BAT sample.

To help place the two samples in a cosmological context, we include a
line in \fig{F:Lx-z} showing the evolution in the position of the knee
(i.e., $L\ast$) of the AGN luminosity function (converted to
10-40~keV\ from the 2-10~keV\ luminosity functions of \citealt{Aird10}
assuming an intrinsic photon index of $\Gamma=1.8$). The position of
the two samples relative to this line demonstrates the complementary
nature of these samples, with the \swift-BAT sample probing the knee
of the luminosity function at $z\sim0.015$ and the ECDFS sample
probing it at $z\sim0.8-1$.  Crucially, this means that due to the
evolution of the SMBH accretion rate density, the power of \nustar\
and \swift-BAT combined will enable us to probe roughly 25\% of the
accretion rate density of the Universe with hard X-rays, compared to
just 0.5\% with \swift-BAT alone.\footnote{These percentages were
  calculated by integrating the evolving X-ray luminosity function of
  \cite{Aird10}.}

\section{\nustar\ sources without \chandra\ or \xmm\ counterparts}
\label{S:Sam}

One of \nustar 's primary science goals is the characterization of the
sources that make up the hard (i.e., $>10$~keV) X-ray background.  As
such, any sources that have not previously been identified at softer
X-ray energies are potentially of great interest.  In this subsection,
we focus on the three \nustar\ sources in the ECDFS field that we have
identified as having neither \chandra\ nor \xmm\ counterparts.  By the
definition of our detection threshold we should expect, on average,
one false source per band.  As such, it is plausible that all three of
these \nustar\ sources with neither \chandra\ nor \xmm\ counterparts
are spurious.  That said, it is important that we first determine
whether any of these three show any other evidence of nuclear activity
before rejecting them as spurious as, if confirmed, they may provide
further insight into the population of Compton-thick AGNs at high
redshift.

The three \nustar\ sources without \chandra\ or XMM-{\it Newton}
counterparts are NuSTAR J033122-2743.9, NuSTAR J033228-2803.5 and
NuSTAR J033231-2737.3 (hereafter, referred to by their indices in our
source catalogue, i.e., 3, 31, and 37, respectively).  These three
sources represent $\sim6\%$ of the total sample of the 49 \nustar\
sources that are significant post-deblending.  We checked by eye the
(5 and 10-pixel Gaussian-smoothed) 2-8~keV \chandra\ images of the
ECDFS near the positions of these three \nustar\ sources, but find no
indication of any weak sources that may have been missed by the source
detection algorithm used in analyzing the \chandra\ data (see L05).
We note that two of the three sources is significantly detected
  in only one of our three \nustar\ bands: one in the 8-24~keV band
  (ID: 3) and one in the 3-24~keV band (ID: 31).  The third (ID: 37)
  is detected in both the 3-8~keV and 3-24~keV bands. Furthermore,
two of these sources have $>3\sigma$ deblended fluxes in at least one
band (which coincides with the band in which it is significantly
detected), the exception being ID 37.  Due to their faintness, all but
one of these three new X-ray sources have poorly constrained photon
indices, the exception being ID: 31, for which BEHR gives a (median)
band ratio of $0.87^{+0.38}_{-0.45}$, corresponding to a photon index
of $\Gamma=1.3^{+1.0}_{-0.3}$.

We next explore whether any of the three new X-ray sources have
counterparts displaying signs of nuclear activity at wavelengths aside
from X-rays (see Figure \ref{F:Thumbs} for 90\arcsec$\times$90\arcsec\
postage-stamp images of the regions around the \nustar\ sources in the
{\it K}-band, {\it Spitzer}-IRAC Ch. 4 and MIPS-24~\mum\ bands and
{\it Herschel}-PACS 100~\mum\ bands).  We first consider the near to
mid-infrared regime (specifically {\it Spitzer}'s IRAC channels), as
this has been shown to be efficient at identifying highly obscured
AGNs (e.g., \citealt{Lacy04, Stern05, Donley07, Donley12}).
Furthermore, considering that $\approx94\%$ of \chandra\ sources in
the ECDFS have counterparts in at least one of the {\it Spitzer}-IRAC
channels (based on a 3\arcsec\ matching to the SIMPLE catalog
described in \citealt{Damen11}), it is highly likely that any genuine
\nustar\ source will also have a {\it Spitzer}-IRAC counterpart.
Indeed, with an on-sky source density of $\sim50$ IRAC sources per
arcmin$^2$, the challenge becomes identifying which, if any, of the
$\sim40$ IRAC sources within our 30\arcsec\ search radius around the
\nustar\ positions is the true counterpart (see Figure
\ref{F:Thumbs}).  To address this, we explore whether any of the
potential counterparts display a rising power-law distribution of IRAC
fluxes, which is evidence of AGN-heated dust (e.g., \citealt{Donley07,
  Donley12}).  Considering the IRAC color criteria of \cite{Donley12},
which are optimized for deep IRAC data unlike the shallow-data
criteria of \cite{Lacy04} and \cite{Stern05}, none of the 95 potential
IRAC counterparts show clear evidence of nuclear activity based on
IRAC data.

\begin{figure}
  \plotone{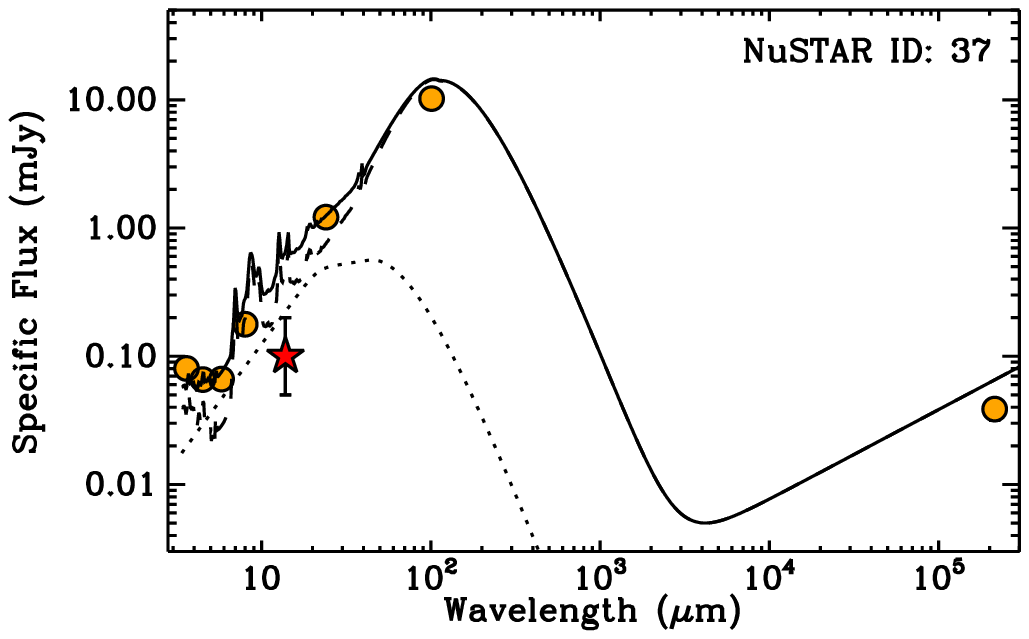}
  \caption{Infrared spectral energy distribution for \nustar -ID: 37.
    Orange circles show data from {\it Spitzer}-IRAC, MIPS, {\it
      Herschel}-PACS and radio data from \cite{Bonzini12}.  On fitting
    the infrared spectral energy distribution of this source using the
    AGN (dotted line) and host-galaxy (solid line) templates described
    in \cite{DelMoro13}, we find that an AGN component is required at
    a significance of $\gg99$\%.  The red star shows the mid-infrared
    flux derived from the \nustar\ 3-8~keV flux using Eqn. 2 of
    \cite{Gandhi09}, which is broadly consistent with the AGN
    component required by the SED fit.}
  \label{F:SED}
\end{figure}

As IRAC power-law selection will identify only those AGN whose
mid-infrared emission strongly dominates over that from
star-formation, we also exploit mid to far-infrared SED fitting to
identify potential infrared signatures of AGN.  To get broader
wavelength coverage for our SED fitting we match to {\it Spitzer}-MIPS
(i.e., the FIDEL catalog; P.I.: M. Dickinson) and {\it Herschel}-PACS
data (i.e., the PEP catalog; \citealt{Lutz11, Magnelli13}), matching
to the positions of the potential {\it Spitzer}-IRAC counterparts
identified above with a search radius of 3\arcsec.  Of the 95
potential IRAC counterparts, only two have detections at 24~\mum\ and
at either 100~\mum\ or 160~\mum\ (our infrared SED fitting require at
least one detection in either of the {\it Herschel}-PACS bands).  The
two potential far-infrared counterparts correspond to \nustar\ sources
31 and 37 (FIDEL catalog IDs: 1294 and 14000; PEP source names:
  PEPPRI~J033229.3-280329 and PEPPRI~J033231.2-273707; and redshifts
  of $z=0.204$ and 0.125 from the MUSYC catalog described in
  \citealt{Cardamone10}, respectively), while \nustar\ source 3 does
not have a significant far-infrared counterpart.  Following
\cite{Mullaney11} we fit the infrared SEDs of these two potential
counterparts, using the extended AGN and host galaxy templates of
\cite{DelMoro13}.  While the counterpart to \nustar\ source 31 shows
no evidence of an AGN component, the counterpart to \nustar\ source 37
requires a significant AGN contribution to fit the infrared SED
(required at $\gg99\%$ significance; see Figure \ref{F:SED}).
Furthermore, converting the \nustar\ 3-8~keV flux to a 2-10~keV
luminosity (assuming $\Gamma=1.8$ and $z=0.125$), and using Eqn. 2
from \cite{Gandhi09} gives a predicted mid-infrared flux broadly
consistent with the AGN contribution required by the SED fit, further
strengthening the case that this is a genuine AGN.  Interestingly, the
SED fit also provides an explanation as to why the {\it Spitzer}-IRAC
fluxes do not display a power-law distribution, with Figure
\ref{F:SED} showing that the SED is likely dominated by emission from
the host galaxy at all infrared wavelengths (e.g.,
\citealt{Cardamone08}).

\begin{figure*}
  \plotone{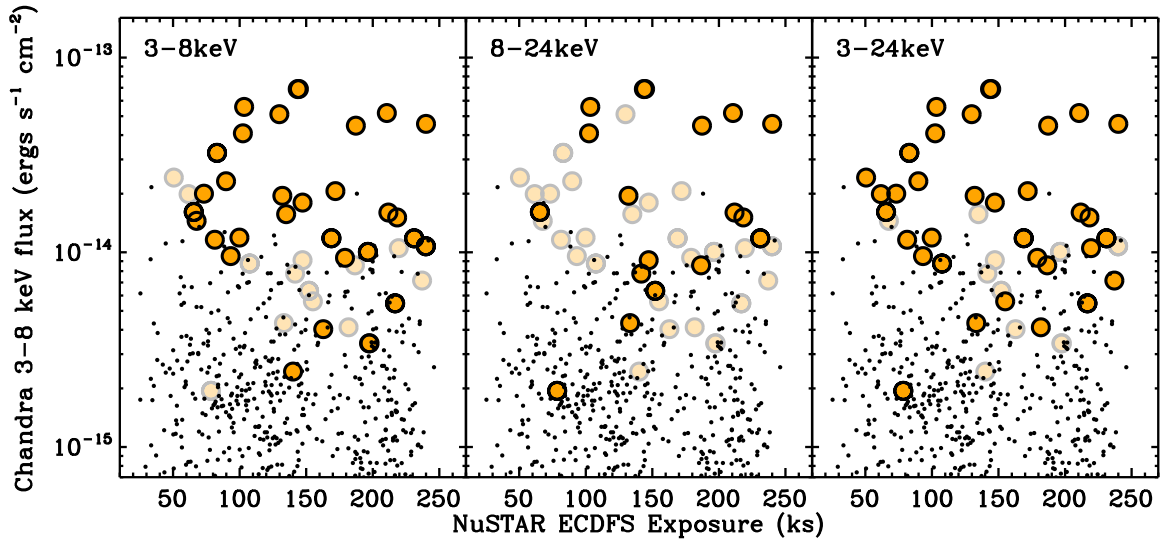}
  \caption{\chandra-derived 3-8~keV fluxes for sources in the
    L05 catalog plotted against the \nustar\ exposure
    time at those positions (small black points).  In each panel, we
    highlight those sources that are detected in the 3-8~keV, 8-24~keV
    and 3-24~keV bands (orange points).  Lighter points show sources
    that are detected by \nustar\ but not in that specific band.  As
    expected, \nustar\ typically detects the brightest sources at a
    given exposure time, although there are examples of undetected
    sources interspersed between the detected sources, suggesting that
    \nustar\ detection is not a simple function of source flux and
    exposure. Reasons for this are explored in \S\ref{S:The}}
  \label{F:Cha}
\end{figure*}

Finally, following \cite{DelMoro13}, we search for potential radio
counterparts to the three \nustar\ sources without \chandra\ or
XMM-{\it Newton} matches.  Matching to the \cite{Bonzini12} catalog
of radio sources in the ECDFS using a 3\arcsec\ search radius around
the IRAC positions, we find that the two potential far-infrared
counterparts identified above are also detected at 1.4~GHz (i.e.,
matched to \nustar\ sources 31 and 37).  As these sources are detected
at far-infrared wavelengths, we explore whether either show a radio
excess above that predicted by the far-infrared/radio correlation that
could indicate the presence of an obscured AGN.  However, we find that
the radio fluxes are consistent with star-formation in both cases (see
Figure \ref{F:SED} for the infrared to radio SED of \nustar\ ID: 37) .

To summarize this section, we detect three \nustar\ sources that have
neither \chandra\ nor XMM-{\it Newton} counterparts.  From the
definition of our detection threshold we expect 2-3 of these to be
spurious (i.e., due to random noise fluctuations).  Analysis of the
infrared SEDs of potential counterparts reveals some evidence that one
of these \nustar\ sources (ID: 37) displays excess flux at
mid-infrared wavelengths that may be attributable to an obscured AGN.
However, no evidence of this AGN is seen at near-infrared wavelengths
(i.e., via power-law {\it Spitzer}-IRAC fluxes) or radio frequencies
(i.e., via a radio excess).  Further evidence of an obscured AGN in
these systems may be identified via other techniques, such as via
their rest-frame optical spectra (i.e., BPT diagnostics;
\citealt{Baldwin81}), although we note that the most heavily obscured
can still be missed using these methods (e.g., \citealt{Stern14}).
However, at present, such spectra are are unavailable for the potential
counterparts identified here.

\section{The \nustar -detected population in the context of the
  \chandra -ECDFS source population}
\label{S:The}

As we have seen, the majority of our significantly detected \nustar\
sources have \chandra\ counterparts.  In this section, we consider how
the \nustar -detected population relates to the wider population of
X-ray sources previously detected in the \chandra\ ECDFS survey.
First, to explore whether the \chandra\ flux is a strong predictor
of whether a source is detected by \nustar\ (as would be expected), we
first plot the distribution of \chandra\ 3-8~keV fluxes for each
source in the L05 catalog against the mean \nustar\ exposure time at
that position, highlighting those that are detected in each of our
three \nustar\ bands (i.e., 3-8~keV, 8-24~keV, 3-24~keV; see Figure
\ref{F:Cha}).  This plot indeed shows that the brightest \chandra\
sources are preferentially detected by \nustar, with 24 of the 34
sources (i.e., $71\%$) with \chandra-measured $F_{\rm
  3-8~keV}>10^{-14}~{\rm ergs~s^{-1}~cm^{-2}}$ and \nustar\ exposure
$>50~{\rm ks}$ detected in at least one \nustar\ band.  Of the other
ten, four seem to be associated with local minima in the $P_{\rm
  False}$ maps that lie just above our detection threshold (i.e.,
having ${\rm log}(P_{\rm False})>-4.5$, compared to our threshold of
$\sim-5.2$).  Thus, while formally undetected, there is at least some
evidence of their presence in the \nustar\ maps.  By contrast, there
is no hint of a reduced $P_{\rm False}$ in the \nustar\ maps at the
positions of the other six bright \chandra\ sources.  Five of these
lie in regions of the \nustar\ maps with comparatively low exposure
(i.e., $<100$~ks), which may explain their non-detection.  One,
however, lies in a region of relatively high exposure (i.e.,
$\sim190$~ks; L05-ID: 577) and thus, with a \chandra\ 3-8~keV flux of
$2\times10^{-14}~{\rm ergs~s^{-1}~cm^{-2}}$ and a photon index of
1.67, we would expect it to be detected by \nustar.  One possibility
for the lack of \nustar\ detection is source variability, with a
factor of a $\sim$few reduction in flux being sufficient to explain
the lack of detection.  However, with similar fluxes between the
\chandra\ and XMM-{\it Newton} observations (separated by $\sim8$
years), this source shows little evidence of strong variability.  As
such, at present it is not clear why the bright \chandra\ source L05:
577 is not detected by \nustar.

Comparing between the individual bands in Figure \ref{F:Cha}, it is
evident that the relationship between the \chandra\ 3-8~keV flux and
\nustar\ detection differs between the \nustar\ bands.  As expected,
detection in the \nustar\ 3-8~keV band is strongly related to the
\chandra\ 3-8~keV flux, with the vast majority of the detections
(i.e., 25 of 33) in this band having \chandra -measured $F_{\rm
  3-8~keV}>10^{-14}~{\rm ergs~s^{-1}~cm^{-2}}$.  However, this
correspondence between detection and \chandra\ flux is weaker in the
8-24~keV band, with only 11 of the 37 \chandra\ sources with \chandra
-measured $F_{\rm 3-8~keV}>10^{-14}~{\rm ergs~s^{-1}~cm^{-2}}$ being
detected in this harder band.  Interestingly, the source with the
faintest \chandra\ 3-8~keV counterpart of all the \nustar\ sources is
detected in the 8-24~keV \nustar\ band, but not the 3-8~keV band
(NuSTAR J033144-2803.0;  NuSTAR-ECDFS ID: 8; L05 ID: 145).  A \nustar\
detection in the 8-24~keV band, but not the 3-8~keV band, indicates a
hard X-ray photon index, but with a \chandra -measured photon index of
$\Gamma=1.51$ for this source, this does not appear to be the case.
Again, we checked for possible counterparts in the later XMM-{\it
  Newton} observations of the field, but found none (i.e., it is
undetected in the XMM-{\it Newton} 2-10~keV band), so we are unable to
rule out source variability as an explanation for the \nustar\
detection of this comparatively faint \chandra\ source.  We do note,
however, that the next faintest \chandra\ source detected in the
\nustar\ 8-24~keV band does have a very hard \chandra -derived photon
index of $\Gamma=-0.72$ (NuSTAR J033145-2745.2; NuSTAR-ECDFS ID: 9;
L05 ID: 152), which would explain its strong detection in this band.

\begin{figure}
  \plotone{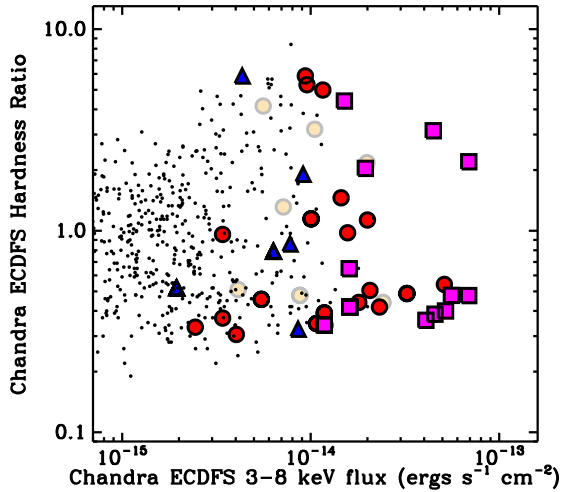}
  \caption{Plot of \chandra -derived 2-8~keV/0.5-2~keV hardness ratio
    against \chandra -measured 3-8~keV flux for detected \chandra\
    sources in the ECDFS (small points).  Highlighted are those
    sources that are detected by \nustar\ in the 3-8~keV band but not
    the 8-24~keV band (red circles), those that are detected in the
    8-24~keV band but not the 3-8~keV band (blue triangles), those
    that are detected in both the 3-8~keV and 8-24~keV bands (magenta
    squares), and those that are only detected in the single 3-24~keV
    band (faint circles).  For those \nustar\ sources with multiple
    \chandra\ counterparts we plot the hardness ratio derived from
    summing the 2-8~keV to 0.5-2~keV band counts from all counterparts
    and the sum of their 3-8~keV fluxes.  While there is a clear
    tendency for the \nustar -detected sources tend to be the
    brightest \chandra\ sources, as expected, there is not clear
    connection between \chandra -derived hardness ratio and whether a
    source is detected by \nustar\ at 8-24~keV.}
  \label{F:Gam}
\end{figure}

In light of this faint, but very hard photon index \chandra\ source
being detected with \nustar , we explore what bearing the \chandra
-derived hardness ratio (i.e., the 2-8~keV to 0.5-2~keV ratio) has on
\nustar\ detections in general.  For this, we plot the \chandra
-derived hardness ratio against \chandra -derived 3-8~keV fluxes for
sources detected in the ECDFS survey (L05), highlighting those that
are detected by \nustar\ (see Figure \ref{F:Gam}).  Here, we only plot
sources with $>50$~ks of \nustar\ coverage.  For those \nustar\
  sources with multiple \chandra\ counterparts we plot the hardness
  ratio derived from summing the 2-8~keV to 0.5-2~keV band counts from
  all counterparts and the sum of their 3-8~keV fluxes.  From this
plot, there is no strong evidence that sources detected in the
\nustar\ 8-24~keV band are preferentially those with harder X-ray
spectra (based on \chandra\ hardness ratios).  Indeed, the majority
(i.e., five of eight) of sources detected in \nustar\ 8-24~keV band
but not the 3-8~keV band have a hardness ratio {\it below} the median
value for all \chandra\ ECDFS sources (i.e., 0.85).  This also holds
for {\it all} \nustar\ sources detected in the 8-24~keV band, not just
those that are not also detected at 3-8~keV (i.e., 15 of 22 have
hardness ratios below the median value).  As such, based on this
analysis, we find no strong evidence that the \chandra\ hardness ratio
has a strong bearing on whether a source is detected by \nustar,
although we acknowledge our relatively small sample size.  Source
variability may be able to explain some of this effect, although
another possibility is that simple \chandra\ hardness ratios are a
poor proxy for the true spectral properties of the X-ray sources
detected by \nustar .  Ascertaining whether this is indeed the case
will be the focus of future papers exploring the spectral properties
of these \nustar\ sources in detail.

\section{Summary}
\label{S:Dis}
In this paper we have described the \nustar\ survey of the ECDFS and
presented the first results from considering the population of sources
detected by \nustar\ in this field.  Such blank field surveys are
  a crucial first step in determining the number counts (i.e.,
  logN-logS) and luminosity function of the resolved hard X-ray (i.e.,
  $>8~{\rm keV}$) source population, which will be the focus of two
  future papers (Harrison et al. in prep. and Aird et al. in
  prep., respectively).  With a maximum unvignetted exposure of
$\approx360$~ks, the ECDFS currently represents the deepest contiguous
component of the \nustar\ extragalactic survey, reaching sensitivity
limits of $\approx1.3\times10^{-14}~{\rm ergs~s^{-1}~cm^{-2}}$
(3-8~keV), $\approx3.4\times10^{-14}~{\rm ergs~s^{-1}~cm^{-2}}$
(8-24~keV) and $\approx3.0\times10^{-14}~{\rm ergs~s^{-1}~cm^{-2}}$
(3-24~keV).  Our main results from this first paper looking at the
population of \nustar\ sources detected in the ECDFS can be summarized
as:

\begin{itemize}
\item within the full $\sim30\arcmin\times30\arcmin$ ECDFS, we detect
  54 \nustar\ sources, although only 49 of these remain significant
  after the effects of neighboring sources are taken into account
  (i.e., after deblending).  Nineteen of these 49 are detected in the
  8-24~keV band - the energy range uniquely probed by \nustar\ to
  unprecedented depths (see \S\ref{SS:Bas});

\item twelve sources are detected in both the 3-8~keV and 8-24~keV
  energy bands, enabling us to determine their hardness ratios and,
  correspondingly, estimate their photon indices within this energy
  range.  The twelve sources display photon indices that span a broad
  range of values, i.e., $\Gamma=0.5 - 2.3$, with a median of
  $\Gamma=1.70\pm0.52$.  By adopting a Monte Carlo approach to sample
  the band ratio probability distribution function of all 49 detected
  sources, we calculate the median photon index for the whole sample
  as $\Gamma=1.90\pm0.53$ (see \S\ref{SS:Bas});

\item all but three of the 49 significantly detected \nustar\ sources
  have X-ray counterparts in either the \chandra\ or \xmm\ surveys of
  the ECDFS.  The redshift of these counterparts span the range
  $z=0.21 - 2.7$ which, when combined with their \nustar\ fluxes,
  corresponds to a 10-40~keV luminosity range for our deblended sample
  of $(\approx0.7 - 300)\times10^{43}~{\rm ergs~s^{-1}}$.  As such, our
  \nustar\ sample probes below the knee of the extrapolated X-ray
  luminosity function at $z<1$ (see \S\S\ref{SS:Bas}, \ref{SS:Com});

\item of the three \nustar\ sources without \chandra\ or \xmm\
  counterparts, only one shows evidence of AGN signatures at other
  wavelengths (i.e., via infrared SED fitting; see \S\ref{S:Sam}); and

\item as expected, a high \chandra\ 3-8~keV flux is a good predictor
  as to whether a source will be detected in other \nustar\ bands.
  However, it is not a 1:1 correlation, with some of the brightest
  \chandra\ sources not being detected at higher energies with \nustar
  .  The reason for this is currently unclear, as the \chandra -
  derived hardness ratio seems to have very little bearing on whether
  a sources is detected in a given \nustar\ band (see \S\ref{S:The}),
  but source variability may provide some explanation.

\end{itemize}

\vspace{3mm}
\noindent
We thank the anonymous referee for their careful reading of the
manuscript and comments which improved the clarity of the text.  This
work made use of data from the \nustar\ mission, a project led by the
California Institute of Technology, managed by the Jet Propulsion
Laboratory, and funded by the National Aeronautics and Space
Administration. We thank the \nustar\ Operations, Software and
Calibration teams for support with the execution and analysis of these
observations. This research has made use of the \nustar\ Data Analysis
Software (NUSTARDAS) jointly developed by the ASI Science Data Center
(ASDC, Italy) and the California Institute of Technology (USA).  ADM
and DMA gratefully acknowledge financial support from the Science and
Technology Facilities Council (ST/I001573/1).  JA acknowledges support
from a COFUND Junior Research Fellowship from the Institute of
Advanced Study, Durham University, and ERC Advanced Grant FEEDBACK at
the University of Cambridge.  FMC acknowledges support from NASA
grants 11-ADAP11-0218 and GO3-14150C.  DRB is supported in part by NSF
award AST 1008067.  WNB and BL thank NuSTAR grant 44A-1092750, NASA
ADP grant NNX10AC99G, and the V. M. Willaman Endowment.  We
acknowledge support from CONICYT-Chile grants Basal-CATA PFB-06/2007
(FEB), FONDECYT 1141218 (FEB) and 1120061 (ET), and "EMBIGGEN" Anillo
ACT1101 (FEB, ET); the Ministry of Economy, Development, and Tourism's
Millennium Science Initiative through grant IC120009, awarded to The
Millennium Institute of Astrophysics, MAS (FEB).  Support for the work
of ET was also provided by the Center of Excellence in Astrophysics
and Associated Technologies (PFB 06).  AC, SP and LZ acknowledge
support from the ASI/INAF grant I/037/12/0– 011/13.  AC acknowledges
the Caltech Kingsley visitor program.  M.\,B. acknowledges support
from NASA Headquarters under the NASA Earth and Space Science
Fellowship Program, grant NNX14AQ07H.

%\bibliography{ecdfs_paper}

\appendix
\section{Catalog Description}
\label{S:Cat}
\begin{table*}
\begin{center}
  \caption{Overview of Columns in the \nustar\ ECDFS Source Catalog}\label{T:Cat}
  \begin{footnotesize} \begin{tabular}{@{}ll@{}} \hline \hline
    Column&\multicolumn{1}{c}{Description}\\
    \hline
    1&Unique NuSTAR ECDFS survey source identification number\\
    2&Name of NuSTAR source\\
    3, 4&Right ascention and declination of the NuSTAR source\\
    5-7&Flag indicating in which of the three standard bands the source is detected\\
    8-10&Flag indicating in which of the three standard bands the
    source is detected post deblending.\\
    11-14&Logarithm of the undeblended false probability in the three standard bands\\
    15-16&Logarithm of the deblended false probability in the three standard bands\\
    17&Flag indicating whether the source remains significant post-deblending\\
    18-32&Total, background and net source counts and associated errors in the three standard bands\\
    33-44&Deblended total, background and net sources counts and associated errors in the three standard bands\\
    45-47&Effective exposure times for the three standard bands\\
    48-62&Total, background and net source count rates and associated errors in the three standard bands\\
    63-68&Deblended total, background and net source count rates and associated errors in the three standard bands\\
    69-73&Band ratio; mean, median, mode and upper and lower limits returned by BEHR algorithm\\
    74-76&Effective photon index and upper and lower limits\\
    77-82&Derived fluxes in the three standard bands\\
    83-88&Derived deblended fluxes in the three standard bands\\
    89&Counterpart catalogue code (L05: \citealt{Lehmer05}; R13:
    \citealt{Ranalli13}, X11: \citealt{Xue11})\\
    90&Unique identification number of the associated source in the counterpart catalogue\\
    91, 92&Right ascention and declination of the associated in the counterpart catalogue\\
    93&Separation between the NuSTAR position and the associated
    source in the counterpart catalogue\\
    94&3-8~keV \chandra\ or XMM-{\it Newton} flux of the associated source in the counterpart catalogue\\
    95&Combined 3-8~keV or \chandra\ or XMM-{\it Newton} flux of all associated sources within 30\arcsec\ of the NuSTAR position\\
    96&Spectroscopic redshift of the associated source in the counterpart catalogue\\
    97&Photometric redshift of the associated source in the counterpart catalogue\\
    98&Adopted redshift\\

    % 95&Absorption-corrected rest-frame 2-10~keV luminosity of the
    % associated source in the counterpart catalogue\\
    % 96&Combined absorption-corrected 2-10~keV luminosity of all associated sources within 30\arcsec\ of the NuSTAR position\\
    99, 100&Non-absorption-corrected rest-frame 10-40keV luminosity and associated error of the \nustar\ source\\
    101&Source notes.\\
    \hline 
  \end{tabular} \end{footnotesize}

\end{center}
\end{table*}
The final catalog of 54 sources generated using the procedure
outlined above is available in the form of an electronic table.  A
summary of the columns of this table is given in Table \ref{T:Cat},
which we expand upon here:

\vspace{1mm}
\noindent
{\it Column 1:} The unique \nustar\ ECDFS identification number.
Sources are arranged in right-ascention order.  Multiple rows labelled
with the same identifier correspond to where one \nustar\ source is
matched to more than one \chandra\ source within 30\arcsec .  

{\it Column 2:} The unique name of the \nustar\ source, following the
IAU-approved naming convention for NuSTAR sources: NuSTAR
JHHMMSS$\pm$DDMM.m, where m is the truncated fraction of an arcminute
in declination for the arcseconds component.

\vspace{1mm}
\noindent {\it Columns 3,4:} The right ascention and declination of
the \nustar\ source.  The position is that of the pixel with the lowest
$P_{\rm False}$ in the 20\arcsec -smoothed maps across all bands.  The
position is taken from the band showing the lowest $P_{\rm False}$
(i.e., the band in which the source is most significantly detected).
All following photometry etc. is then performed at this one position
in all three bands.

\vspace{1mm}
\noindent {\it Columns 5-7:} Binary flag indicating in which of the
three standard bands the source is formally detected (i.e., meets our
final thresholds of ${\rm log}(P_{\rm False}) \leq -5.19, \leq -5.22$
and $\leq -5.34$ in the 3-8~keV, 8-24~keV and 3-24~keV bands,
respectively).  Here, and throughout the table, the 3-8~keV, 8-24~keV
and 3-24~keV bands are abbreviated to SB, HB and FB, respectively.

\vspace{1mm}
\noindent {\it Columns 8-10:} The same as columns 5-7, but after photon
counts from neighboring sources have been excluded (i.e., deblending;
see \S\ref{SSS:Net}).

\vspace{1mm}
\noindent
{\it Columns 11-13:} The logarithm of $P_{\rm False}$ (i.e., ${\rm
  log}(P_{\rm False})$) in each of the three standard bands based on a
20\arcsec\ smoothing length, i.e., the smoothing length adopted to
assess whether a source is significantly detected (see
\S\ref{SSS:Bli}).

\vspace{1mm}
\noindent
{\it Columns 14-16:} The same as columns 11-13, but after
photon counts from neighboring sources have been excluded (i.e.,
deblending; see \S\ref{SSS:Net}).

\vspace{1mm}
\noindent
{\it Column 17:} Binary flag indicating whether the \nustar\ source
remains significant in at least one of the standard bands
post-deblending.

\vspace{1mm}
\noindent {\it Columns 18-32:} Non-aperture-corrected total,
background and net source counts based on 30\arcsec\ aperture
photometry centered on the position in columns 3,4 in each of the three
standard bands.  Errors are given on the total and net counts and are
calculated using \cite{Gehrels86} (see \S\ref{SSS:Net}).  We indicate
those sources where the photometric measurement in a given band is
$<3\sigma$ with a negative value for the error.

\vspace{1mm}
\noindent
{\it Columns 33-44:} Same as columns 18-32, but after source counts
from neighboring sources have been taken into account (still based on
30\arcsec\ aperture photometry).  Errors are given on the net counts.

\vspace{1mm}
\noindent
{\it Columns 45-47:} Mean effective exposure times in 30\arcsec\
apertures centered on the position in columns 3,4 in each of the three
standard bands. Unit: s.

\vspace{1mm}
\noindent {\it Columns 48-62:} Non-aperture-corrected total,
background and net source count rates, calculated from the source
counts in columns 18-32 and the respective mean effective exposure
times in columns 45-47).  Errors are propagated from those on the
source counts, assuming zero error on the effective exposure times.
Again, negative error values indicate those sources that are detected
at $<3\sigma$ in a given band. Unit: ${\rm s^{-1}}$.

\vspace{1mm}
\noindent {\it Columns 63-68:} Same as columns 48-62, but after the
effects of source blending has been taken into account (i.e.,
calculated using the source counts given in columns 33-44 and
respective mean effective exposure times given in columns 45-47).
Unit: ${\rm s^{-1}}$.

\vspace{1mm}
\noindent {\it Columns 69-73:} 8-24~keV to 3-8~keV band ratios output
by the BEHR algorithm.  As this algorithm is a Bayesian Estimator, it
calculates the band ratio probability distribution function and
provides the mean, median, mode and upper and lower 68th percentiles,
which we report here.

\vspace{1mm}
\noindent {\it Columns 74-76:} Effective observed (i.e.,
non-absorption-corrected) photon index of the \nustar\ source derived
from the mean band ratio and the upper and lower limits given in
columns 69-73 (see \S\ref{SSS:Flu}).

\vspace{1mm}
\noindent {\it Columns 77-82:} Observed-frame, aperture corrected
source fluxes in the three standard bands.  Fluxes are derived from
net count rates (columns 48-62) following the procedure outlined in
\S\ref{SSS:Flu}.  Errors are propagated using the same conversion
factors.  Again, negative error values indicate those sources that are
detected at $<3\sigma$ in a given band.  Units: ${\rm
  ergs~s^{-1}~cm^{-2}}$

\vspace{1mm}
\noindent {\it Columns 83-88:} Same as columns 77-82, but after the
effects of source blending have been taken into account (i.e., using the
deblended count-rates in columns 63-68).

\vspace{1mm}
\noindent {\it Columns 89:} Code indicating the catalog in which a
\chandra\ or XMM-{\it Newton} counterpart to the \nustar\ source was
identified.  L05: \cite{Lehmer05}; R13: \cite{Ranalli13}; X11: \cite{Xue11}.

\vspace{1mm}
\noindent {\it Column 90:} unique identifier of the matched \chandra\
\chandra\ or XMM-{\it Newton} source(s) from the catalog indicated in
column 89 (``recno'' in the L05 and X11 catalogs; ``ID210'' in the R13
catalog) .  Where there are more than one match within 30\arcsec\ of
the \nustar\ position, we provide separate matches on additional table
rows.  Such multiple matches can therefore be identified by the
replication of the \nustar\ index in column 1.

\vspace{1mm}
\noindent {\it Columns 91,92:} Right-ascention and declination of the
respective \chandra\ or XMM-{\it Newton} match(es).

\vspace{1mm}
\noindent {\it Column 93:} Separation between the \nustar\ source and
the matched \chandra\ or XMM-{\it Newton} source. Unit: arcsecond.

\vspace{1mm}
\noindent {\it Column 94:} Observed-frame 3-8~keV fluxes derived from
the \chandra\ or XMM-{\it Newton} data.  Note: these data are {\it
  not} provided in the catalogs of L05, \cite{Ranalli13}
or \cite{Xue11} as the 3-8~keV band is not a standard band of those
papers.  Instead, the 3-8~keV fluxes were calculated using ACIS
Extract (\citealt{Broos10}) with a simple power-law model. Units:
${\rm ergs~s^{-1}~cm^{-2}}$.

\vspace{1mm}
\noindent {\it Column 95:} Total combined observed-frame 3-8~keV
fluxes of {\it all} \chandra\ or XMM-{\it Newton} sources within
30\arcsec\ of the \nustar\ source. Units: ${\rm ergs~s^{-1}~cm^{-2}}$.

\vspace{1mm}
\noindent {\it Column 96:} Spectroscopic redshift of the matched
\chandra\ (from \citealt{Silverman10} ) or XMM-{\it Newton} (from
\citealt{Ranalli13}) source, where available.

\vspace{1mm}
\noindent {\it Column 97:} Photometric redshift of the matched
\chandra\ (from \citealt{Silverman10} ) or XMM-{\it Newton} (from
\citealt{Ranalli13}) source, where available.

\vspace{1mm}
\noindent {\it Column 98:} The adopted redshift used to calculate the
rest-frame 10-40~keV luminosity of the \nustar\ source.  In cases where
both spectroscopic and photometric redshifts are available, we always
adopt the spectroscopic redshift.

\vspace{1mm}
\noindent {\it Columns 99, 100:} Derived rest-frame 10-40~keV
luminosities of the \nustar\ sources and associated errors for sources
where a redshift is available (either photometric of spectroscopic).
Luminosities are calculated using the observed 8-24~keV band flux and
$k$-corrections are performed assuming a photon index of $\Gamma=1.8$
(see \S\ref{SSS:Mul}).  Where a \nustar\ source is matched to more
than one \chandra\ or XMM-{\it Newton} source (and therefore may have
more than one associated redshift), we provide the 10-40~keV
luminosity assuming each of the various available redshifts.  Errors
are propagated directly from the errors on the 8-24~keV fluxes (i.e.,
we assume zero error on the redshift).  Again, negative error values
indicate those sources that are detected at $<3\sigma$ in a given
band. Units: ${\rm ergs~s^{-1}}$

\vspace{1mm}
\noindent {\it Column 101:} Any special notes associated with the source.

\end{document}